\documentclass[twocolumn,floatfix,tightenlines,superscriptaddress]{revtex4}
\usepackage{mathtools}
\usepackage{bm}
\usepackage{dsfont,amsthm,amsbsy}
\usepackage{verbatim}
\usepackage{amssymb}
\usepackage{amsmath}
\usepackage{bbm}
\usepackage{graphicx}
\usepackage{epstopdf}
\usepackage{subfigure}
\usepackage{natbib}
\usepackage{epsfig}
\usepackage{amsfonts}
\usepackage{mathrsfs}
\usepackage{sidecap}
\usepackage{lipsum}
\usepackage[toc,page,title,titletoc,header]{appendix}
\usepackage[colorlinks,linkcolor=blue,citecolor=blue,anchorcolor=blue,urlcolor=blue]{hyperref}
\usepackage{resizegather}
\usepackage{tikz}
\usepackage{float}
\usepackage{mathbbol}
\usepackage[normalem]{ulem}
\usepackage{cancel}
\usepackage{upgreek}
\usepackage{dblfloatfix}
\usepackage{dcolumn}
\usepackage{xcolor}
\usepackage{braket}
\usepackage{pdflscape} 
\usepackage{float}

\begin{document}
	
	\title{Scarred discrete time crystal in a periodically driven dimerized spin chain}
	\author{Davood Marripour}
	\author{Saeed S. Jahromi}
   \author{Jahanfar Abouie}
	\email[Corresponding author: ]{jahan@iasbs.ac.ir}

	\affiliation{Department of Physics, Institute for Advanced Studies in Basic Sciences (IASBS), Zanjan 45137-66731, Iran}

	\date{\today}
	
	\begin{abstract}
We investigate the emergence of a scarred discrete time crystal (SDTC) phase in a periodically driven dimerized spin chain. While generic interacting Floquet systems are expected to thermalize according to the eigenstate thermalization hypothesis (ETH), we demonstrate that this system hosts quantum many-body scars (QMBS) that induce a regime of weak ergodicity breaking. Through an analysis of Floquet level statistics, entanglement entropy, and eigenstate fidelity, we identify a manifold of low-entanglement states characterized by semi-Poisson statistics embedded within an otherwise thermal spectrum. These scarred states support robust subharmonic oscillations with period doubling, signaling the spontaneous breaking of discrete time-translation symmetry. We show that the SDTC response is robust against a variety of initial state configurations, demonstrating its stability beyond fine-tuned conditions. A finite-size scaling analysis reveals that the time-crystalline lifetime grows with system size within the range accessible to our exact-diagonalization calculations. However, drawing on the general phenomenology of approximate many-body scars, we expect that hybridization between Floquet scars and the thermal continuum will eventually curtail this growth, causing the lifetime to saturate at system sizes beyond our current numerical reach. This characterizes the SDTC as a long-lived metastable dynamical regime rather than a strictly stable thermodynamic phase, providing a comprehensive framework for understanding the interplay between periodic driving and constrained many-body dynamics in disorder-free systems.
\end{abstract}
	
	\maketitle

\section{Introduction}
\label{sec:introduction}
In recent years, the study of non-equilibrium dynamics in quantum many-body systems has uncovered phenomena that challenge traditional statistical mechanics. According to the eigenstate thermalization hypothesis (ETH) \cite{wang2024ETH,srednicki1994,rigol2008}, generic interacting systems undergoing unitary evolution rapidly lose memory of their initial local information and approach thermal equilibrium. However, ergodicity breaking and ETH violation have been identified in various contexts, including integrable models \cite{dunajski2012}, many-body localized (MBL) systems \cite{basko2006,nandkishore2015}, Hilbert space-constrained models \cite{kumar2024hilbert,Langlett,Moudgalya1}, and prethermal phases \cite{B. Bauer,K. Mallayya,D. V. Else,G. He,S. A. Weidinger,E. Canovi,M. Bukov,Kyprianidis,Das Sarma,Zeng,Stasiuk,Nandkishore,Lazarides,Kjall,huse,Bordia,Oganesyan,Keyserlingk,Johri,Huse,P. Ponte,Igloi2007}.

Among these, quantum many-body scars (QMBS) \cite{kunimi2024,bernien2017,moudgalya2018}, a form of weak ergodicity breaking, have emerged as a compelling area of study. Unlike MBL systems, where ergodicity is broken across the entire energy spectrum, QMBS arise from a subset of atypical, low-entropy eigenstates embedded within a dense thermalizing background \cite{bluvstein2021,su2022}. Experimentally, these states can be probed through specific initial configurations that exhibit persistent quantum revivals, contrasting with the featureless dynamics of thermal states.

Parallel to these developments, Floquet engineering has enabled the realization of novel dynamical phases \cite{oka2019}, most notably discrete time crystals (DTCs) \cite{khemani2016,else2016,marripour2025,Das2026,marripour2026}. These phases break discrete time-translation symmetry, exhibiting a robust subharmonic response to a periodic drive. While stabilization of DTCs was originally thought to necessitate MBL-induced disorder \cite{yao2017,zhang2017}, recent research demonstrates that dynamical constraints in disorder-free systems can also protect long-lived temporal order \cite{Huang2018,H. Yar}. This convergence has led to the study of scarred discrete time crystals (SDTCs) \cite{maskara2021,sugiura2021}, where periodic driving is used to enhance the stability of scarred revivals \cite{bluvstein2021,ho2019,su2022}. A key outstanding question is whether such temporal order can be robustly stabilized without strict kinetic constraints \cite{bernien2017,turner2018,hudomal2022,mukherjee2020}, and how integrability-breaking perturbations govern the eventual "melting" of this phase into the thermal continuum.
\begin{figure}[tpb]
		\centering
		\includegraphics[width=0.95 \columnwidth]{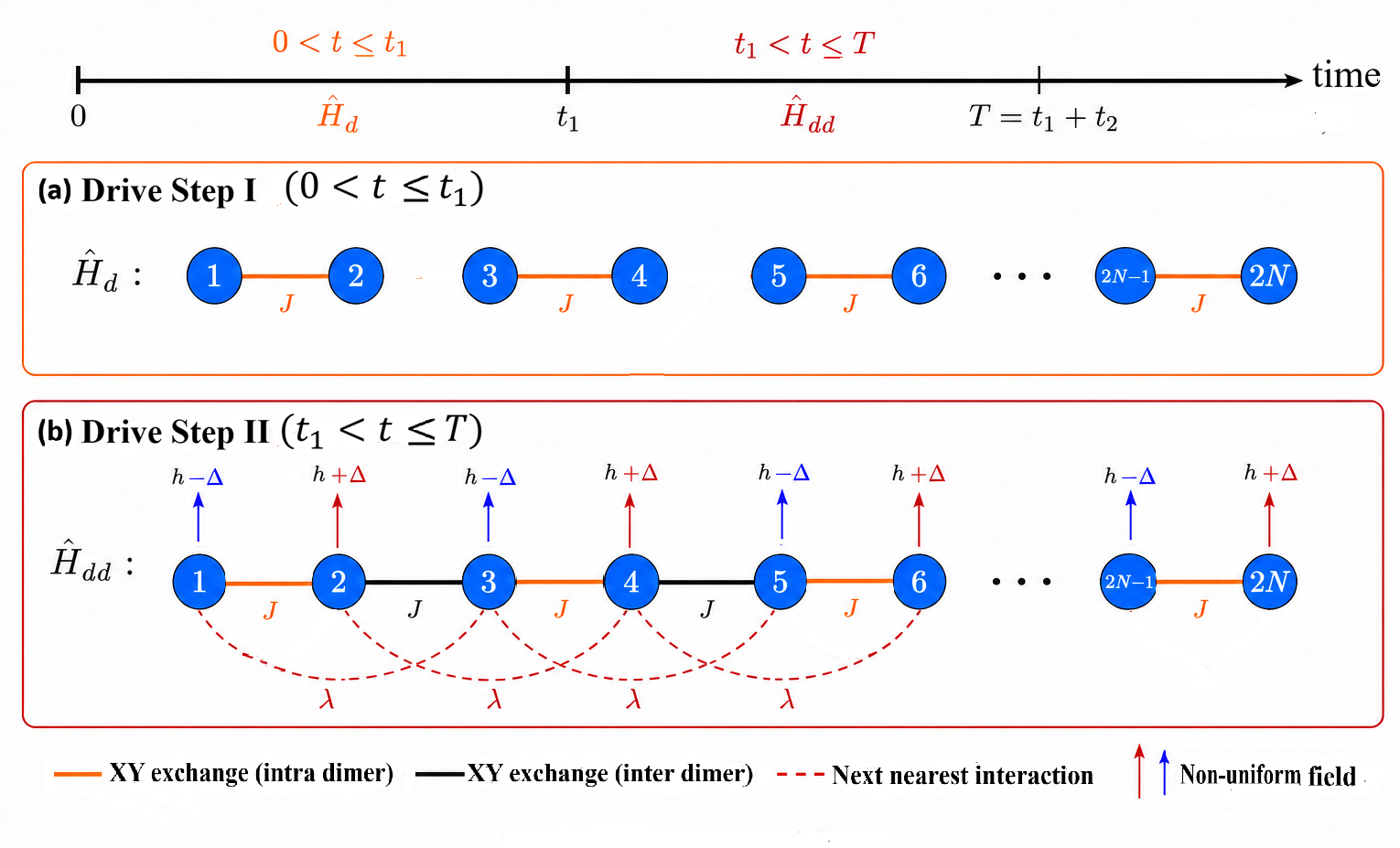}
		\caption{Schematic representation of the periodically driven dimerized spin chain. The driving cycle 
$T=t_1+t_2$ consists of two distinct phases: during $t_1$, the dimers evolve as non-interacting units, while during $t_2$, a dimer-dimer interaction and an external magnetic field are introduced.}
		\label{fig:schematic}
	\end{figure}

In this paper, we investigate the non-equilibrium dynamics of a periodically driven, antiferromagnetic dimerized spin chain (Fig. \ref{fig:schematic}). By introducing an integrability-breaking perturbation $J$, we systematically tune the system from a non-ergodic regime to a fully chaotic thermal phase. Working in the zero-magnetization sector, we identify the presence of QMBS, where the Floquet spectrum exhibits semi-Poisson level statistics ($0.39<\langle r \rangle < 0.53$) and hosts a manifold of low-entanglement outlier states that violate ETH. We demonstrate that these scarred eigenstates possess anomalously large overlaps with the N\'{e}el state, leading to robust, period-doubled subharmonic oscillations that persist despite local quantum fluctuations. Finally, we map the transition as $J$ increases, showing that as the scarred subspace melts, the subharmonic response decays and the level statistics recover Wigner-Dyson (WD) behavior ($\langle r \rangle \approx 0.53$), signaling the restoration of global ergodicity.

The remainder of this paper is organized as follows. Section \ref{sec:model} introduces the dimerized Hamiltonian and the identification of QMBS via spectral statistics. Section \ref{sec:entanglement} investigates the entanglement entropy of Floquet eigenstates, contrasting the thermal bulk with scarred outliers. Section \ref{sec:overlaps} details the overlap structure between Floquet eigenstates and the initial N\'{e}el state, while Sec.~\ref{sec:dynamical} presents the signatures of time-translation symmetry breaking via staggered magnetization and autocorrelation functions. Finally, we conclude in Sec.~\ref{sec:conclusion}.

%--------------

\section{Periodically driven dimerized spin chain}\label{sec:model}

We consider a chain of $N$ non-interacting dimers (comprising $L=2N$ spins), as illustrated in Fig.~\ref{fig:schematic}. Each dimer is described by the XY Hamiltonian:
\begin{equation}
\hat{H}_d = J (\sigma_1^x \sigma_{2}^x + \sigma_1^y \sigma_{2}^y),
\label{Eq:Hd}
\end{equation}
where $J > 0$ is the antiferromagnetic coupling constant and $\sigma^{x,y}$ are the Pauli operators. The ground state is the singlet $|s\rangle = \frac{1}{\sqrt{2}}(|\uparrow\downarrow\rangle - |\downarrow\uparrow\rangle)$ with energy $E_s=-2J$. The excited states consist of the $m_s = \pm 1$ components, $|t^\pm\rangle = \{|\uparrow\uparrow\rangle, |\downarrow\downarrow\rangle\}$, at zero energy, and the $m_s = 0$ state, $|t^0\rangle = \frac{1}{\sqrt{2}}(|\uparrow\downarrow\rangle + |\downarrow\uparrow\rangle)$, at energy $E=2J$. In the uncoupled limit, these excitations are localized to individual dimers; however, weak inter-dimer interactions allow these excitations to acquire kinetic energy and form dispersive bands.

The system is subjected to a Floquet drive consisting of two alternating periods within a cycle $T=t_1+t_2$. During $t_1$, the dimers are governed by $H_d$. During $t_2$, an inter-dimer interaction is activated and the Hamiltonian of the system is given by:
\begin{eqnarray}
\nonumber \hat{H}_{dd}&=&\sum_i \left[ J(\sigma_i^x\sigma_{i+1}^x+\sigma_i^y\sigma_{i+1}^y)+\lambda \sigma_{i-1}^z\sigma_{i+1}^z \right]\\
 &&+\sum_i(h+\Delta (-1)^i)\sigma_i^z,
\label{Eq:Hdd}
\end{eqnarray}
where the sums run over all lattice sites, $\lambda$ represents antiferromagnetic exchange coupling, while $h$ and $\Delta$ denote the uniform and staggered longitudinal magnetic fields, respectively. 

The model's integrability is broken by the coupling $J$. The presence of a staggered field $\Delta$ reduces the translational symmetry of the lattice, explicitly breaking the one-site translation $\mathcal{T}: \sigma_i^{\alpha} \to \sigma_{i+1}^{\alpha}$ (for $\alpha \in \{x, y, z\}$) while preserving the two-site translation $\mathcal{T}^2: i \to i+2$. This reduction to a two-site unit cell establishes the dimerized structure of the system.

Furthermore, the staggered field $\Delta$ breaks both the global $\mathbb{Z}_2$ symmetry ($\sigma_i^z \to -\sigma_i^z$) and the lattice reflection symmetry ($i \to L-i+1$). These combined effects—the symmetry reduction induced by $\Delta$ and the overall non-integrable nature of the time-dependent driving—preclude the existence of protected conservation laws. Consequently, the system is driven into a generic, non-integrable regime, fostering ergodic thermalization.

Under open boundary conditions, the only remaining exact microscopic symmetry is the global $U(1)$ symmetry associated with the conservation of total magnetization $\sigma_{total}^{z}$. This symmetry leads to a fragmentation of the Hilbert space into disjoint sectors, a constraint we account for by restricting our analysis to a specific symmetry sector. Because the microscopic discrete symmetries are broken, any observed subharmonic response cannot be attributed to symmetry-protected mechanisms or eigenstate pairing, making this system an ideal platform to investigate the genuine spontaneous breaking of discrete time-translation symmetry. We focus our investigation on the strong-interaction regime, defined by $\lambda \gtrsim \Delta$ and $J \neq 0$.

\subsection{Level spacing statistics}

To characterize ergodicity and investigate the potential breakdown of thermalization in our periodically driven dimer chain, we analyze the statistical properties of the Floquet spectrum. The long-time dynamics are governed by the unitary time-evolution operator defined over one period $T=t_1 + t_2$:
\begin{equation}
U = e^{-i\hat{H}_{dd} t_2} e^{-i \hat{H}_d t_1} \quad (\hbar=1).
\end{equation}
The physical evolution is determined by the products of the energy parameters and their respective time durations,for instance, $Jt_1$, $Jt_2$, $\lambda t_2$, and $\Delta t_2$. These dimensionless products, or \textit{pulse areas}, dictate the rotation angles of the state vector within each driving interval. Consequently, the Floquet spectrum and the system's long term stability depend on these integrated couplings rather than the energy scales in isolation. The eigenvalues of the evolution operator, $e^{-i \varepsilon_n T}$, yield the quasienergies $\varepsilon_n \in [-\pi/T, \pi/T)$, which serve as the fundamental descriptors of the many-body Floquet spectrum.

To provide microscopic intuition, we consider the evolution of an isolated dimer under $\hat{H}_d$. For an initial state $|\psi(0)\rangle=|\uparrow\downarrow\rangle$, which can be decomposed into the singlet $|s\rangle$ and the $m_s=0$ triplet $|t^0\rangle$ states, the time-evolved state is:
\begin{equation}
|\psi(t)\rangle = \cos(2Jt)|\uparrow\downarrow\rangle - i\sin(2Jt)|\downarrow\uparrow\rangle.
\label{Eq:time_evol}
\end{equation}
This demonstrates coherent oscillations between $|\uparrow\downarrow\rangle$ and $|\downarrow\uparrow\rangle$ with an angular frequency $\omega = 2J$. At $t=t_1$, the probability of the state swap is $P_{\text{swap}}=\sin^2(2Jt_1)$. To ensure a perfect swap at the end of each first interval, we satisfy the condition $2Jt_1=(2n+1)\pi/2$, setting $t_1=\pi/(4J)$ for $n=0$. 

The second interval, $t_2$, is dedicated to the evolution under the inter-dimer interaction Hamiltonian $\hat{H}_{dd}$, which mediates many-body correlations and breaks the microscopic symmetries. To systematically isolate the influence of interaction strength, we factor out $\lambda$ from $\hat{H}_{dd}$ and define a rescaled interacting Hamiltonian $\tilde{H}_{dd}(\tilde{J}, \tilde{h}, \tilde{\Delta})$ with dimensionless parameters $\tilde{J}=J/\lambda$, $\tilde{h}=h/\lambda$, and $\tilde{\Delta}=\Delta/\lambda$. Under this transformation, the unitary evolution during the second interval is governed by the scaled pulse area $\lambda t_2$, such that $U_2 = e^{-i (\lambda t_2) \tilde{H}_{dd}}$.

This rescaling highlights a fundamental invariance in the system's dynamics: for a fixed set of dimensionless parameters $\{\tilde{J}, \tilde{h}, \tilde{\Delta}\}$, the Floquet spectrum and long-time correlations depend solely on the pulse area $\lambda t_2$. Consequently, dimer chains with distinct physical exchange and field parameters will exhibit identical dynamical evolution, provided their scaled parameters and the product $\lambda t_2$ remains invariant. Conversely, varying $\lambda t_2$ effectively modulates the interaction strength relative to the pulse duration drives the system through different regimes of thermalization and ergodicity, offering a clear protocol to map out the phase diagram of the driven chain.

In our specific protocol, we fix $t_2=1$, which renders the interaction pulse area directly proportional to the coupling strength $\lambda$. Because $t_1$ is set to $\pi/(4J)$ to satisfy the local $\pi$-pulse condition, the total driving period $T=1+\pi/(4J)$ becomes $J$-dependent, resulting in a tunable Floquet frequency $\omega_0(J)=2\pi/(1+\pi/(4J))$. This approach decouples the local coherent oscillations required for optimal excitation transfer—from the collective many-body dynamics driven by $\lambda$, allowing us to probe the transition between distinct dynamical regimes as a function of interaction strength.

The proposed protocol is platform-independent. While absolute energy scales vary by orders of magnitude—ranging from the kHz regime in ultracold atomic gases to the THz regime in solid state semiconductor spin systems, the underlying Floquet dynamics are determined by the dimensionless pulse areas (e.g., $\lambda t_2$ and $Jt_1$). Consequently, the predicted many-body phenomena, such as subharmonic responses and scar-protected dynamics, are universal and applicable across a wide range of experimental architectures.

To distinguish between ergodic (thermalizing) and non-ergodic phases, we analyze the spectral fluctuations using two complementary measures: the distribution of normalized level spacings, $P(s)$, and the average level spacing ratio, $\langle r \rangle$. The latter is derived from the ratio of consecutive quasienergy spacings, $r_n$, defined as:
\begin{equation}
    r_n = \frac{\min(\delta_n, \delta_{n-1})}{\max(\delta_n, \delta_{n-1})},
    \label{eq:level_spacing}
\end{equation}
where $\delta_n = \varepsilon_n - \varepsilon_{n-1}$ denotes the spacing between consecutive quasienergies. In the following, $s = \delta_n / \langle \delta \rangle$ represents the spacing normalized by the mean level spacing $\langle \delta \rangle$.

In the ergodic limit, the system is expected to follow WD statistics. For a time-reversal invariant system, such as ours, this corresponds to the Gaussian Orthogonal Ensemble (GOE), where the level spacing distribution is $P(s) = \frac{\pi}{2}s \exp(-\frac{\pi}{4}s^2)$. This chaotic regime is characterized by a high degree of spectral stiffness and an average ratio of $\langle r \rangle \approx 0.53$ \cite{Atas2013}. Conversely, integrable or many-body localized (MBL) systems exhibit a strongly broken ergodicity characterized by Poisson statistics, $P(s) = e^{-s}$, where the lack of level repulsion leads to an average ratio of $\langle r \rangle \approx 0.39$ \cite{huse,turner2018}.

\begin{figure}[ht]
    \centering
    \includegraphics[scale=0.27]{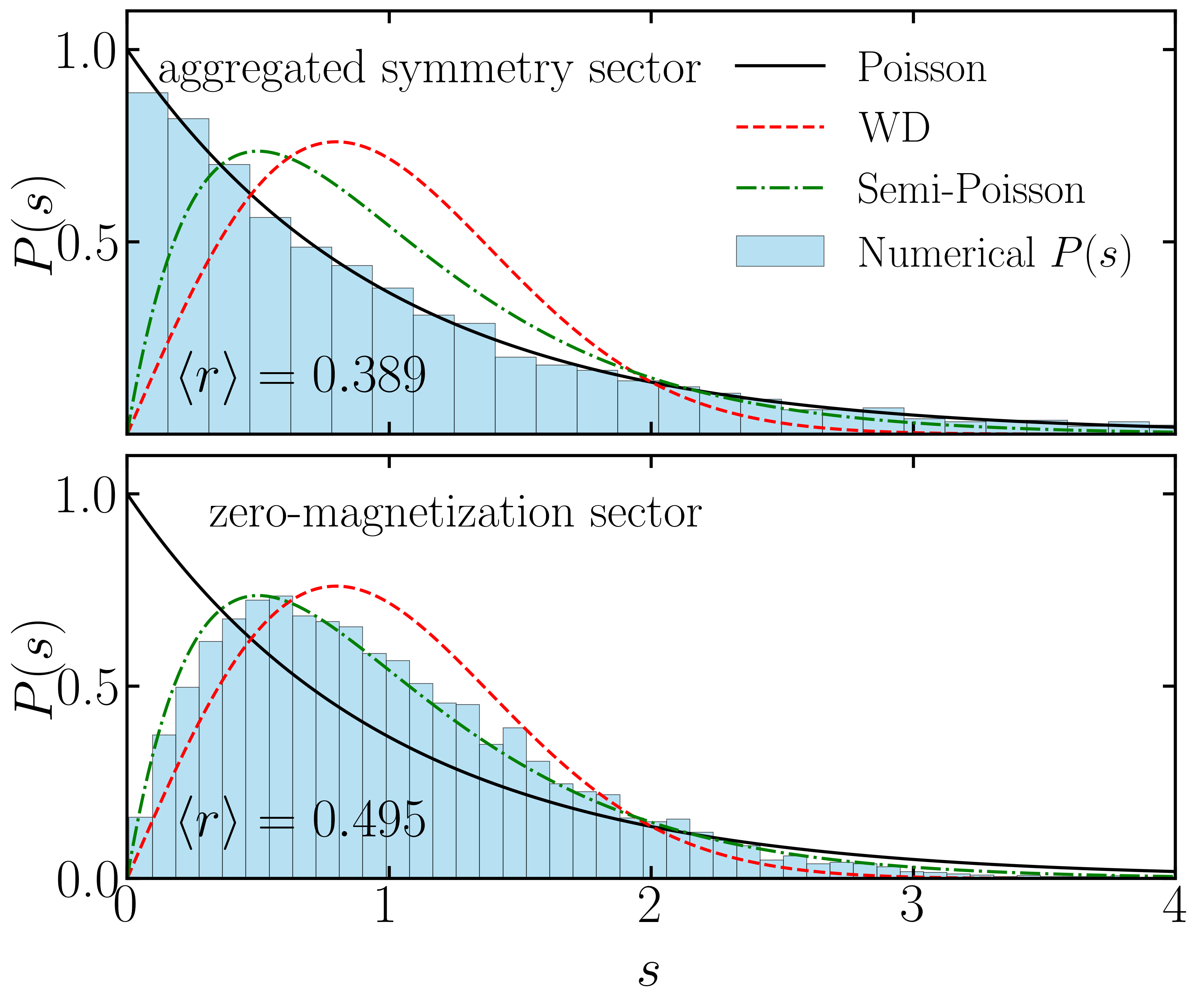}
    \caption{(Top): Level spacing distribution $P(s)$ obtained from the aggregated symmetry sectors. The resulting Poissonian statistics highlight the necessity of partitioning the global $U(1)$ symmetry to avoid spurious non-ergodic signatures. (Bottom): Level spacing distribution $P(s)$ for the driven dimer chain with $\tilde{\Delta}= 0.11$, $\tilde{J}=0.15$, $\tilde{h}=1$ and $\lambda t_2= 0.9$, and $L=16$. The numerical results in the $\{\tilde{J}, \tilde{\Delta}\}<1$ regime exhibit semi-Poisson statistics, signaling constrained dynamics and many-body scars. Theoretical Poisson and WD distributions are shown for comparison.}
    \label{fig:level_spacing}
\end{figure}

The dimer spin model in Eqs.~(\ref{Eq:Hd}) and (\ref{Eq:Hdd}) possesses a global $U(1)$ symmetry, as the Hamiltonians $H_d$ and $H_{dd}$ commute with $\sigma^z_{total} = \sum_i \sigma_i^z$. This symmetry partitions the Hilbert space into independent sectors. A critical aspect of the spectral analysis is the proper partitioning of these sectors; if all symmetry sectors are pooled, the level spacing statistics remain Poissonian regardless of the internal dynamics (see Fig.~\ref{fig:level_spacing}, top). This occurs because energy levels from uncorrelated sectors do not exhibit level repulsion, resulting in a spurious Poissonian signature that mimics non-ergodic behavior. To reveal the true thermalization properties and distinguish between integrable and chaotic regimes, we restrict our analysis to the zero-magnetization sector ($\sigma^z_{total} = 0$), which constitutes the largest symmetry sector.

To explicitly demonstrate these spectral features, we examine the level spacing distribution, $P(s)$ in Fig.~\ref{fig:level_spacing}. As the integrability-breaking parameter $J$ increases, the system moves away from integrable dynamics. However, in the limit of dominant inter-dimer coupling ($\{\tilde{J}, \tilde{\Delta}\}<1$), the competition between the inter-dimer interactions and the local dimer structure prevents the emergence of fully chaotic mixing. Our numerical results reveal a distinct regime of \textit{weakly broken ergodicity}, where the level spacing distribution follows semi-Poisson statistics, $P(s) = 4s e^{-2s}$, with an intermediate average ratio of $\langle r \rangle \approx 0.495$.

This semi-Poissonian behavior occupies a unique middle ground between the two ergodic extremes. The vanishing of $P(s)$ as $s \to 0$ indicates that the system retains a degree of level repulsion, distinguishing it from the strongly non-ergodic Poisson limit. Nevertheless, the exponential decay of the distribution tail, rather than the Gaussian decay characteristic of the GOE signals that spectral stiffness is suppressed. This serves as a hallmark of constrained dynamics, where the underlying dimer geometry imposes effective selection rules that prevent the system from fulfilling the ETH, even in the absence of strong disorder. Such behavior is indicative of a mixed phase space or Hilbert space fragmentation, potentially pointing to the emergence of QMBS induced by the periodic driving \cite{Schmit1999}.

\begin{figure}[ht]
    \centering
    \includegraphics[scale=0.285]{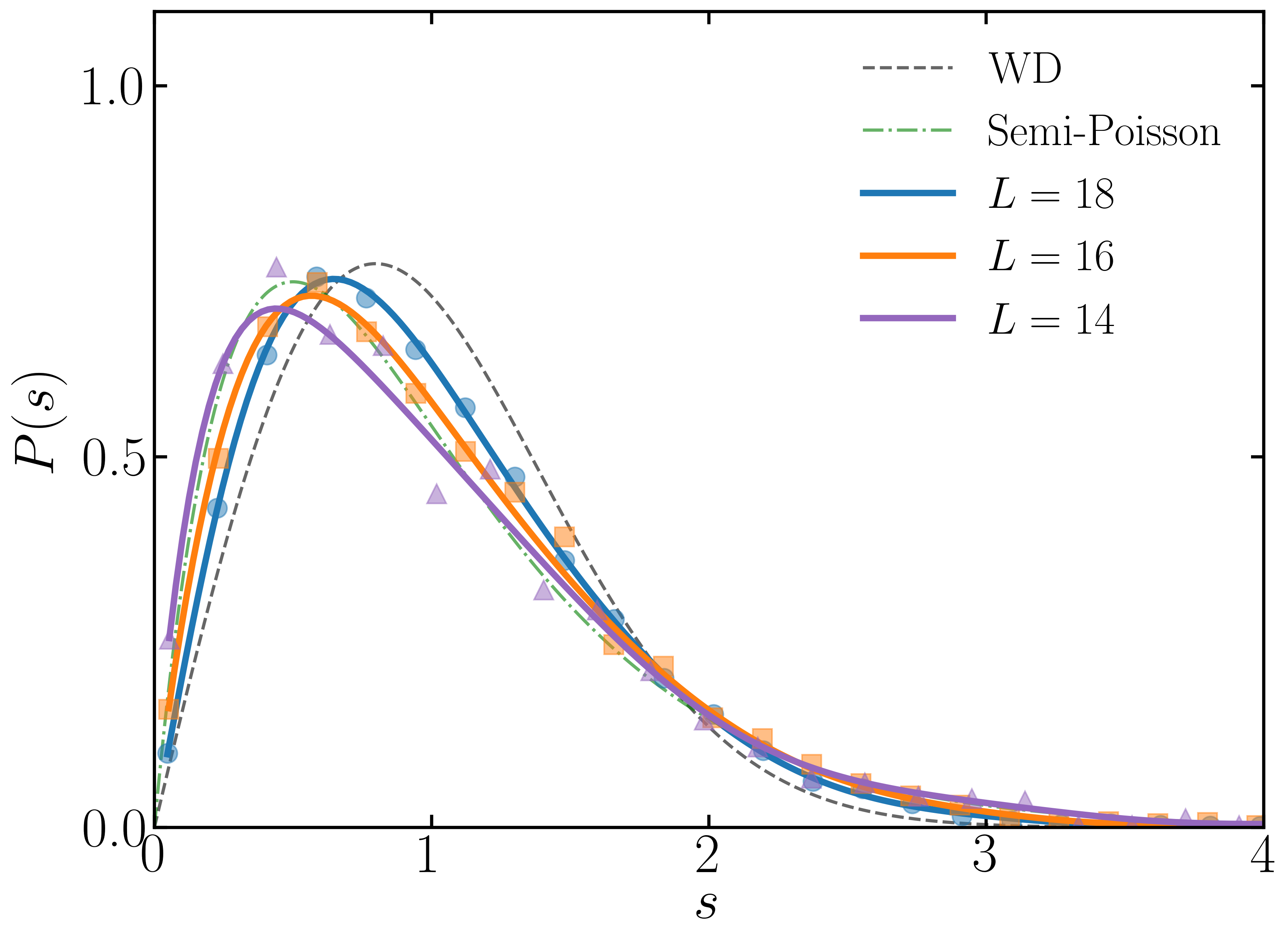}
    \caption{Level spacing distribution $P(s)$ for system sizes $L=14, 16$, and $18$ in the weakly non-ergodic regime. Here, $\tilde{\Delta}= 0.11$, $\tilde{J}= 0.15$, $\tilde{h}=1$ and $\lambda t_2= 0.9$. The green dash-dotted and red dashed curves denote the analytical predictions for semi-Poisson and WD statistics, respectively. The observed finite-size flow indicates that as $L$ increases, the numerical data systematically depart from the semi-Poissonian regime toward the WD distribution, signaling an eventual crossover to thermalization in the thermodynamic limit.}
    \label{fig:ps_size}
\end{figure}

Furthermore, we investigate how this weakly non-ergodic regime evolves under changes to the system size $L$ and the inter-dimer coupling $J$. As illustrated in Fig.~\ref{fig:ps_size}, increasing the system size $L$ induces a systematic shift in the level spacing distribution from the semi-Poisson prediction toward the WD limit. This finite-size flow suggests that the constraints imposed by the dimer geometry are a finite-size effect, and ergodicity is eventually restored in the thermodynamic limit.

\begin{figure}[ht]
    \centering
    \includegraphics[scale=0.28]{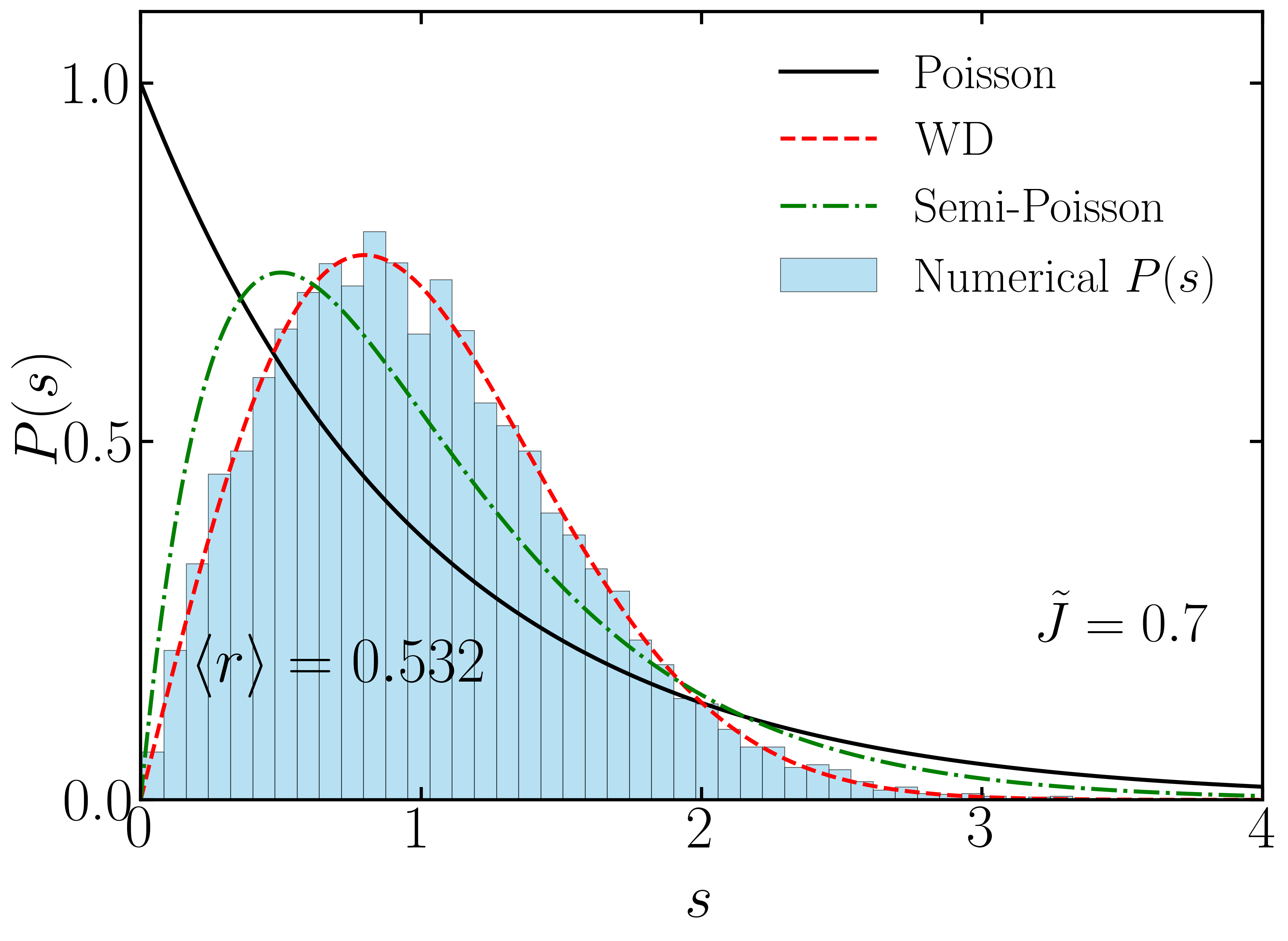}
    \caption{Level spacing distribution $P(s)$ for the periodically driven chain of length $L=16$, with $\tilde{\Delta}= 0.11$, $\tilde{h}=1$ and $\lambda t_2=0.9$, evaluated at the larger coupling $\tilde{J}=0.7$. At this value of $\tilde{J}$ the statistics shift from the semi-Poissonian regime toward the WD distribution, indicating a transition into the fully ergodic thermal phase.}
    \label{fig:level_spacing2}
\end{figure}

Complementary to the finite-size scaling, we examine the effect of increasing the integrability-breaking strength $J$. As shown in Fig.~\ref{fig:level_spacing2}, for a fixed system size ($L=16$), a significant increase in $J$ drives the spectral statistics away from the semi-Poissonian regime and toward the WD ensemble. This transition marks the crossover from the constrained, weakly non-ergodic phase into a fully ergodic thermal phase, where the system obeys the ETH. A more exhaustive treatment of the spectral statistics and the methodology employed is provided in Appendix~\ref{app:level_statistics}.

\begin{figure}[ht]
    \centering
    \includegraphics[scale=0.25]{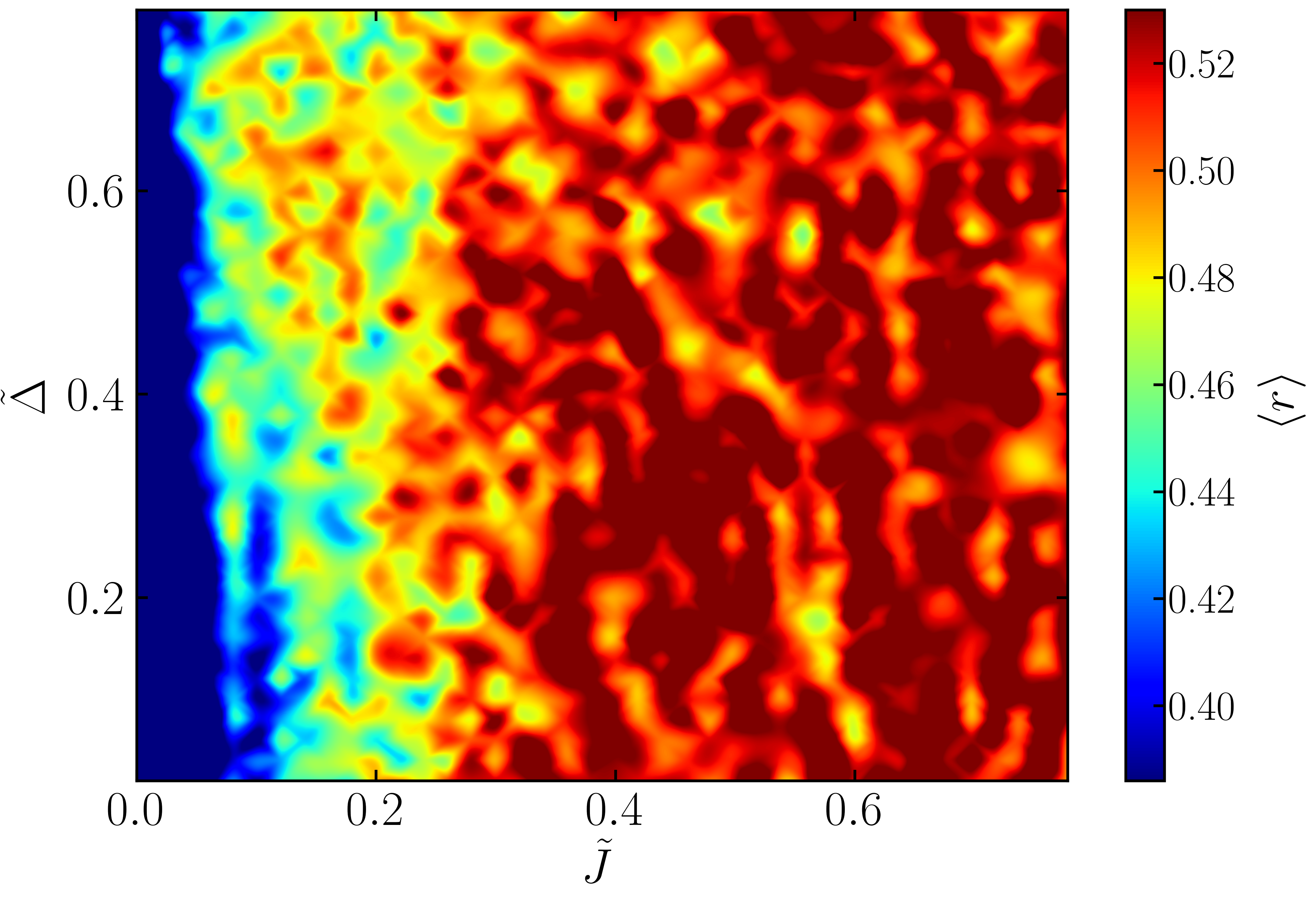}
    \caption{Heat map of the average level-spacing ratio $\langle r \rangle$ in the $\tilde{J}-\tilde{\Delta}$ plane for $L=14$ and $\lambda t_2=0.9$. The color scale spans from the Poisson limit ($\langle r \rangle \approx 0.39$, dark blue) to the Wigner-Dyson limit ($\langle r \rangle \approx 0.53$, dark red). The diagram highlights a transition from a constrained, non-thermal regime at low $\tilde{J}$ to a fully ergodic thermal phase at higher $\tilde{J}$, with the transition being largely insensitive to the detuning $\tilde{\Delta}$.}
    \label{fig:map}	
\end{figure}

To delineate the boundaries between the non-thermal and thermal phases, we construct a phase diagram of the average level spacing ratio $\langle r \rangle$ in the $\tilde{J}-\tilde{\Delta}$ plane (Fig.~\ref{fig:map}). The heat map reveals a distinct transition in the system's spectral statistics. For weak inter-dimer coupling ($\tilde{J} \lesssim 0.22$), the system resides in a non-thermal regime characterized by intermediate $\langle r \rangle$ values, consistent with the semi-Poissonian statistics discussed previously. As $\tilde{J}$ increases beyond this threshold, the system undergoes a crossover into a globally ergodic thermal phase where $\langle r \rangle$ approaches the WD limit of $\approx 0.53$. Notably, the phase boundary is predominantly vertical, suggesting that while the coupling $\tilde{J}$ is the primary driver of integrability breaking, the detuning parameter $\tilde{\Delta}$ plays a comparatively secondary role in determining the ergodicity of the system.

\begin{figure}[ht]
    \centering
    \includegraphics[scale=0.28]{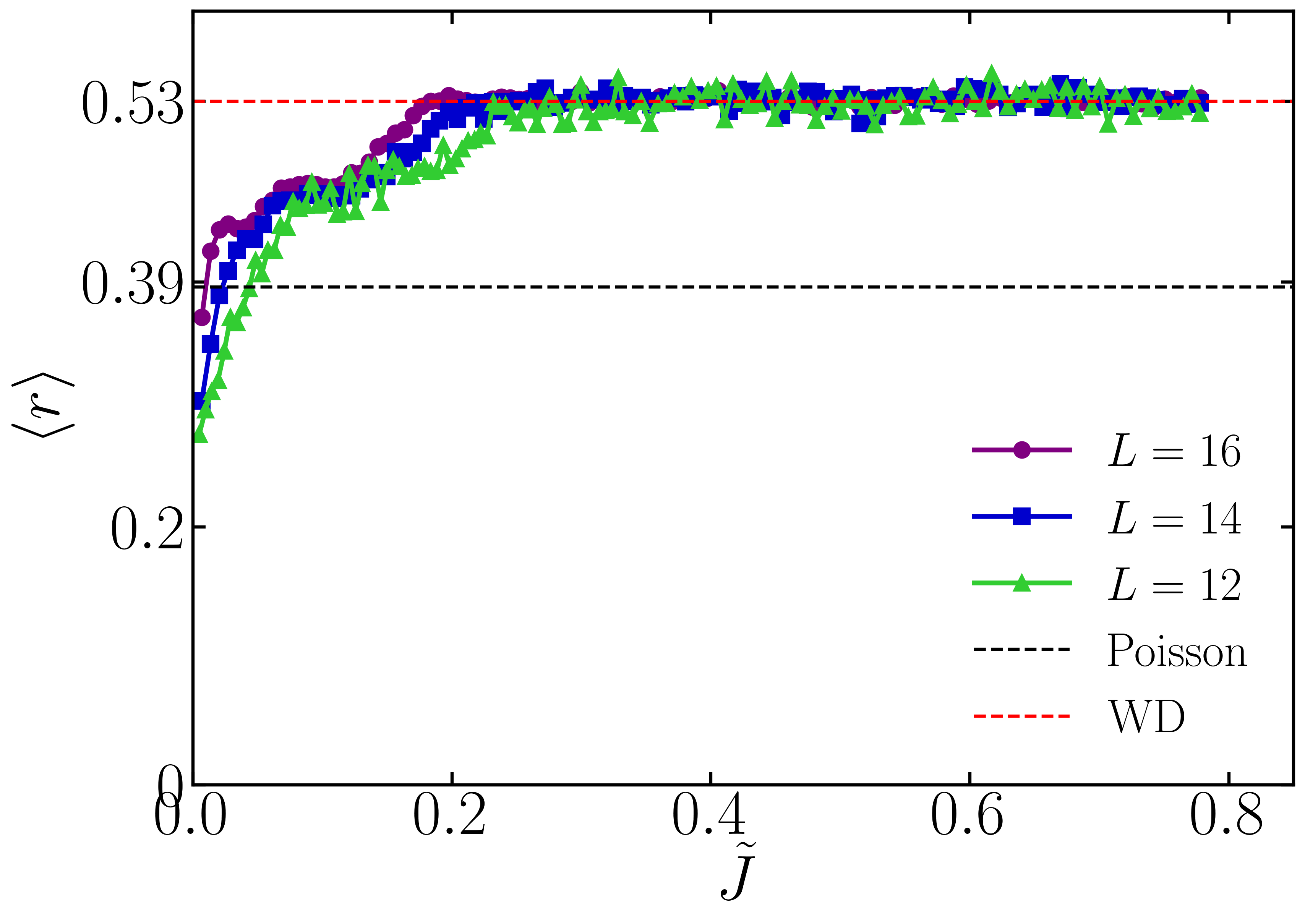}
    \caption{Average level-spacing ratio $\langle r \rangle$ versus the coupling $\tilde{J}$ for system sizes $L=12, 14$, and $16$, with $\tilde{\Delta}= 0.11$ and $\lambda t_2=0.9$. The black and red dashed lines denote the theoretical Poisson ($\approx 0.39$) and WD ($\approx 0.53$) limits, respectively. The increasing sharpness of the crossover with $L$ is a signature of the transition from the constrained regime to a fully ergodic thermal phase.}
    \label{fig:R_Vs_J}	
\end{figure}

We further investigate the scaling properties of this transition by plotting $\langle r \rangle$ as a function of $\tilde{J}$ for various system sizes $L$ (Fig.~\ref{fig:R_Vs_J}). In the limit of vanishing coupling ($\tilde{J}\to 0$), the ratio $\langle r \rangle$ lies significantly below the Poisson limit, indicating a highly constrained or integrable regime. As $\tilde{J}$ increases, the system exhibits a rapid crossover toward the GOE limit. This transition displays clear finite-size scaling behavior: as the system size $L$ increases, the crossover becomes increasingly sharp and shifts toward lower values of $\tilde{J}$. For $\tilde{J} \gtrsim 0.22$ (at $\lambda t_2=0.9, \tilde{\Delta}=0.11$), the ratio saturates at the GOE value of $\approx 0.53$ across all studied sizes. This saturation confirms that the system recovers ergodicity and enters a robust thermalizing phase in the large-$\tilde{J}$ limit, characterized by the level repulsion statistics expected for chaotic quantum many-body systems.

It is important to distinguish the global spectral properties
from the scarred eigenstates. The semi-Poisson statistics reflect the partially
suppressed level repulsion characterizing the entire zero-magnetization sector,
whereas the quantum many-body scars form a sparse set of outlier eigenstates
embedded within this bulk. The global semi-Poisson nature of the spectrum and
the distinct high-fidelity scarred states that retain the N\'eel memory are
therefore two independent signatures of the model's constrained dynamics.

\section{Entanglement entropy and thermalization dynamics}\label{sec:entanglement}

The transition from the non-ergodic to the thermal regime is further substantiated by the entanglement properties of the Floquet eigenstates.

For a given eigenstate $|\psi_n\rangle$, we define the bipartite entanglement entropy (EE) as the von Neumann entropy of the reduced density matrix $\rho_A$:
\begin{equation}
S_n = -\mathrm{Tr}(\rho_A \ln \rho_A),
\end{equation}
where $\rho_A = \mathrm{Tr}_B \left(|\psi_n\rangle\langle\psi_n|\right)$ is the reduced density matrix of subsystem $A$, obtained by tracing out the degrees of freedom of the complementary subsystem $B$ (with $L_A = L/2$). In our numerical computations, the half-chain bipartition is chosen such that each subsystem contains an equal number of spins. Specifically, for an even number of dimers, the cut is placed at an inter-dimer bond, whereas for an odd number of dimers, it is placed at an intra-dimer bond. In either case, the cut crosses three links: two $\lambda$-type links and one additional inter- or intra-dimer bond. Since the inter- and intra-dimer exchange couplings are taken to be equal ($J$), these bipartitions yield identical physical results; for example, the half-chain cut is inter-dimer for $L=16$ and intra-dimer for $L=18$, yet both are physically equivalent. All calculations are performed within the zero-magnetization sector, consistent with our spectral analysis.

According to the Floquet-ETH, a generic non-integrable periodically driven system is expected to thermalize to an infinite-temperature ensemble. In this ergodic limit, the EE follows a volume-law scaling and approaches the Page value, $S_{\text{Page}}$, which characterizes the average entanglement of a random pure state in the Hilbert space \cite{Page1993}. For a one-dimensional spin-1/2 chain of length $L$ bipartitioned at $L_A = L/2$:
\begin{equation}
	S_{\text{Page}}=\frac{L\ln2-1}{2}.
\end{equation}

\begin{figure}[t]
    \centering
    \includegraphics[scale=0.3]{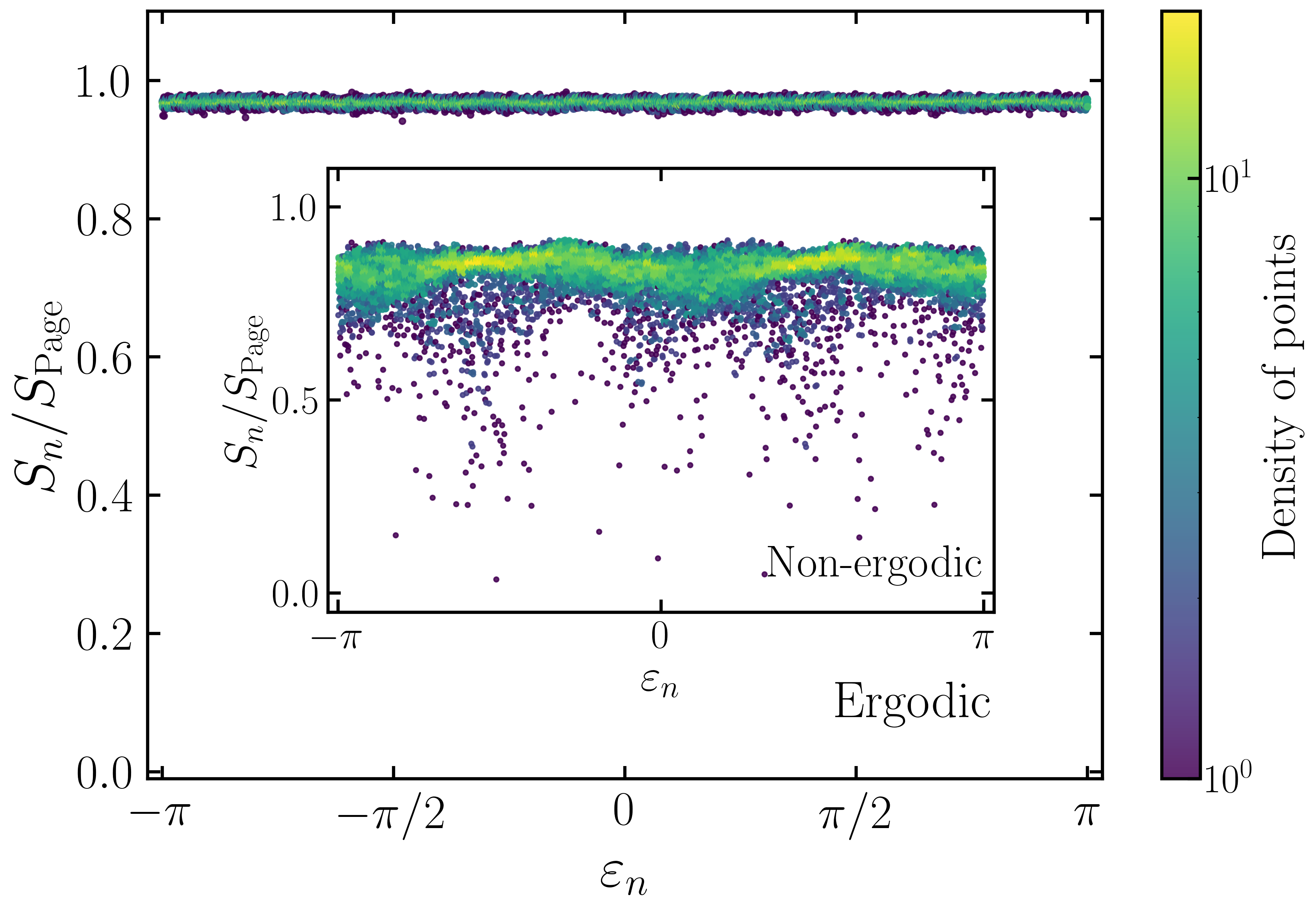}
    \caption{Normalized EE $S_n/S_{\text{Page}}$ versus quasi-energy $\varepsilon_n$ ($L=16$). Main panel: At $\tilde{J}=0.56$, states cluster near the Page value ($S_n/S_{\text{Page}} \approx 1$), consistent with strong Floquet-ETH and the absence of QMBS. Inset: At $\tilde{J}=0.17$, a distinct population of low-EE outliers emerges alongside the thermal bulk, signifying QMBS that violate Floquet-ETH and stabilize SDTC dynamics. Color scales indicate eigenstate density; $\tilde{\Delta}=0.11$ and $\lambda t_2=0.9$ throughout.}
    \label{fig:entropy_thermal}
\end{figure}

To explicitly test this in the ergodic phase, we compute the EE for $\tilde{J}=0.56$, $\tilde{\Delta}=0.11$, and $\lambda t_2=0.9$. Figure~\ref{fig:entropy_thermal} shows the normalized EE, $S_n/S_{\text{Page}}$, as a function of the quasi-energy $\varepsilon_n \in [-\pi, \pi]$ for a chain of size $L=16$. The results reveal that the normalized EE for nearly all Floquet eigenstates is concentrated in a narrow band at $S_n/S_{\text{Page}} \approx 1$. This tight clustering at the Page value demonstrates that the eigenstates are nearly maximally entangled, obeying strict volume-law scaling. The absence of low-entropy outlier states across the entire quasi-energy spectrum provides rigorous evidence that the system satisfies the strong Floquet-ETH. Consequently, any local initial state subject to this driving protocol will act as its own thermal bath, rapidly losing its initial quantum memory and thermalizing to a featureless infinite-temperature state.

In the presence of persistent subharmonic oscillations, the EE exhibits a significant deviation from the standard Floquet-ETH predictions. We identify a regime characterized by Floquet-QMBS: a sparse set of outlier states that, despite occupying a negligible fraction of the Hilbert space, exert a dominant influence on the system's long-time dynamics.

To characterize these states, we examine the EE for a representative configuration in the many-body regime. As shown in Figure~\ref{fig:entropy_thermal} (inset), the normalized EE $S_n/S_{\text{Page}}$ as a function of quasienergy $\varepsilon_n$ reveals a dense, highly entangled thermal bulk adhering to the Floquet-ETH. Superimposed on this bulk is a distinct set of anomalous eigenstates with significantly reduced entanglement. These scarred states constitute the microscopic mechanism that shields specific initial configurations, such as the Néel state, from rapid thermalization, thereby stabilizing the long-lived subharmonic revivals characteristic of the SDTC phase.

\section{Fidelity of Floquet eigenstates and the N\'eel state}\label{sec:overlaps}

\begin{figure}[tbp]
    \centering
    \includegraphics[scale=0.27]{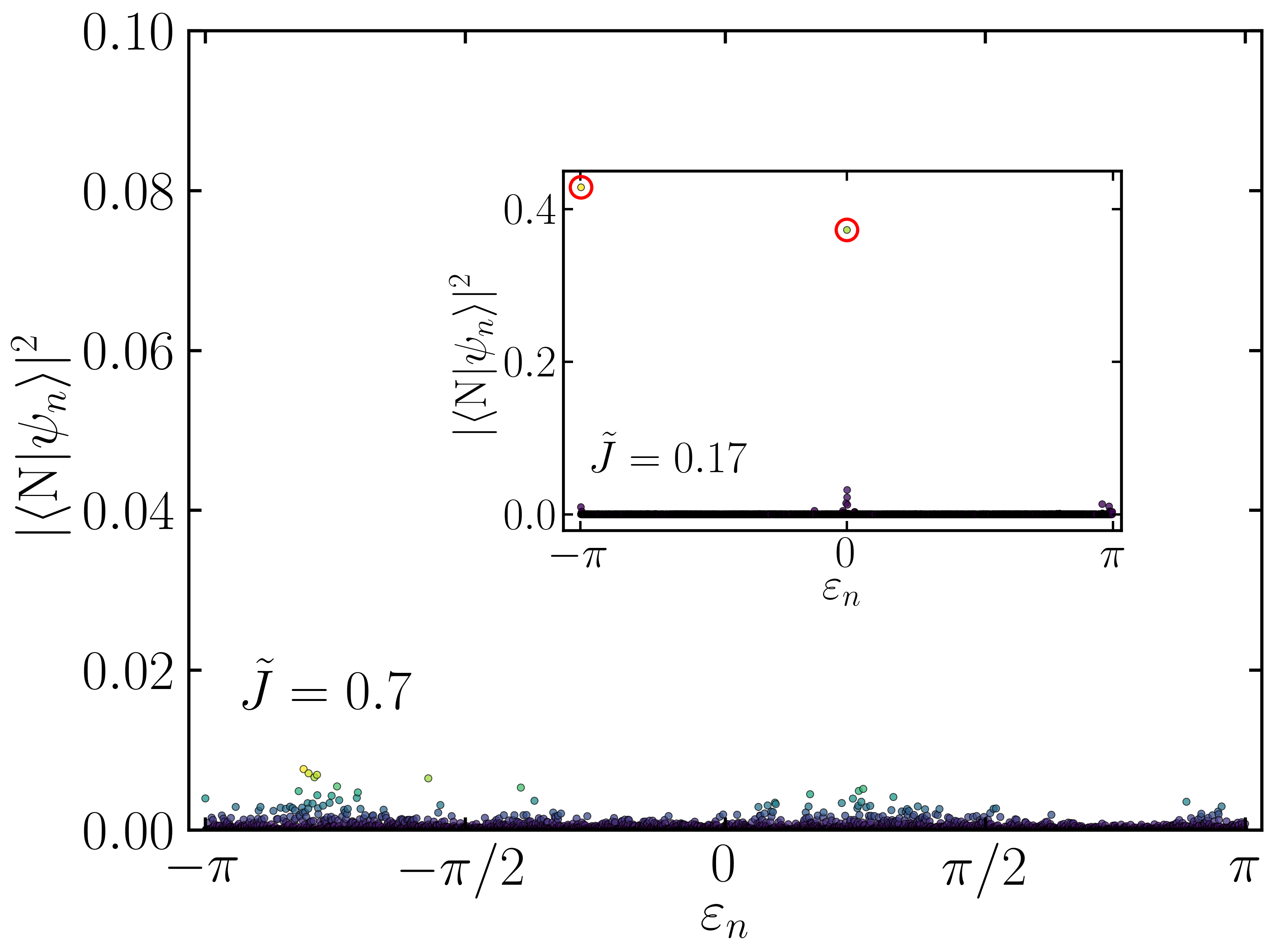}
\caption{Fidelity $|\langle N| \psi_n \rangle|^2$ versus quasi-energy $\varepsilon_n$. In the scarred regime ($\tilde{J}=0.17$), a distinct subset of eigenstates maintains high overlap with the Néel state, indicating QMBS-induced nonthermal behavior. Conversely, in the deep ergodic regime ($\tilde{J}=0.7$), these outliers vanish, and fidelity is uniformly suppressed to $\mathcal{O}(1/\mathcal{D})$, consistent with strong Floquet-ETH and complete thermalization. Parameters: $\tilde{\Delta}=0.11$ and $\lambda t_2=0.9$.}
   \label{fig:overlap}
\end{figure}

The emergence of low-entanglement outliers serves as a key signature of the SDTC phase, which can be further substantiated by examining the overlap (fidelity) between the Floquet eigenstates $|\psi_n\rangle$ and the initial N\'eel state, $|N\rangle = |\uparrow \downarrow \uparrow \downarrow \dots \rangle$. Figure~\ref{fig:overlap} displays the fidelity $|\langle N| \psi_n \rangle|^2$ across the quasi-energy spectrum for both the scarred and ergodic regimes.

In the scarred regime, the system exhibits a striking departure from standard thermalization. While the vast majority of Floquet eigenstates (the thermal bulk) possess exponentially suppressed overlaps with the N\'eel state, scaling as $\mathcal{O}(1/\mathcal{D})$, a rare subset of eigenstates exhibits anomalously high fidelity. These high-overlap states correspond precisely to the low-entanglement outliers identified in the previous section. By retaining an extensive memory of the N\'eel configuration, these non-thermal states provide the microscopic mechanism that underpins the long-lived, subharmonic revivals characteristic of the SDTC phase.

Conversely, increasing the driving parameter, drives the system deep into the ergodic phase, where integrability breaking and state mixing processes dominate. In this regime, the high-fidelity states vanish entirely, and the overlap with the N\'eel state becomes uniformly suppressed across the entire spectrum. This disappearance of scar states signals the crossover to purely ergodic dynamics, confirming that all initial states, including the N\'eel state, undergo rapid, featureless thermalization.

\section{The scarred discrete time crystalline (SDTC) phase}\label{sec:dynamical}

Having established the existence of a scarred subspace and its anomalous overlap with the N\'eel state, we now characterize the direct dynamical signatures of the DTC phase. The hallmark of a DTC is the spontaneous breaking of the discrete time-translational symmetry imposed by the Floquet drive, manifesting as a robust subharmonic response with a period $2T$.

\begin{figure}[htbp]
	\centering
	\includegraphics[width=0.48 \textwidth]{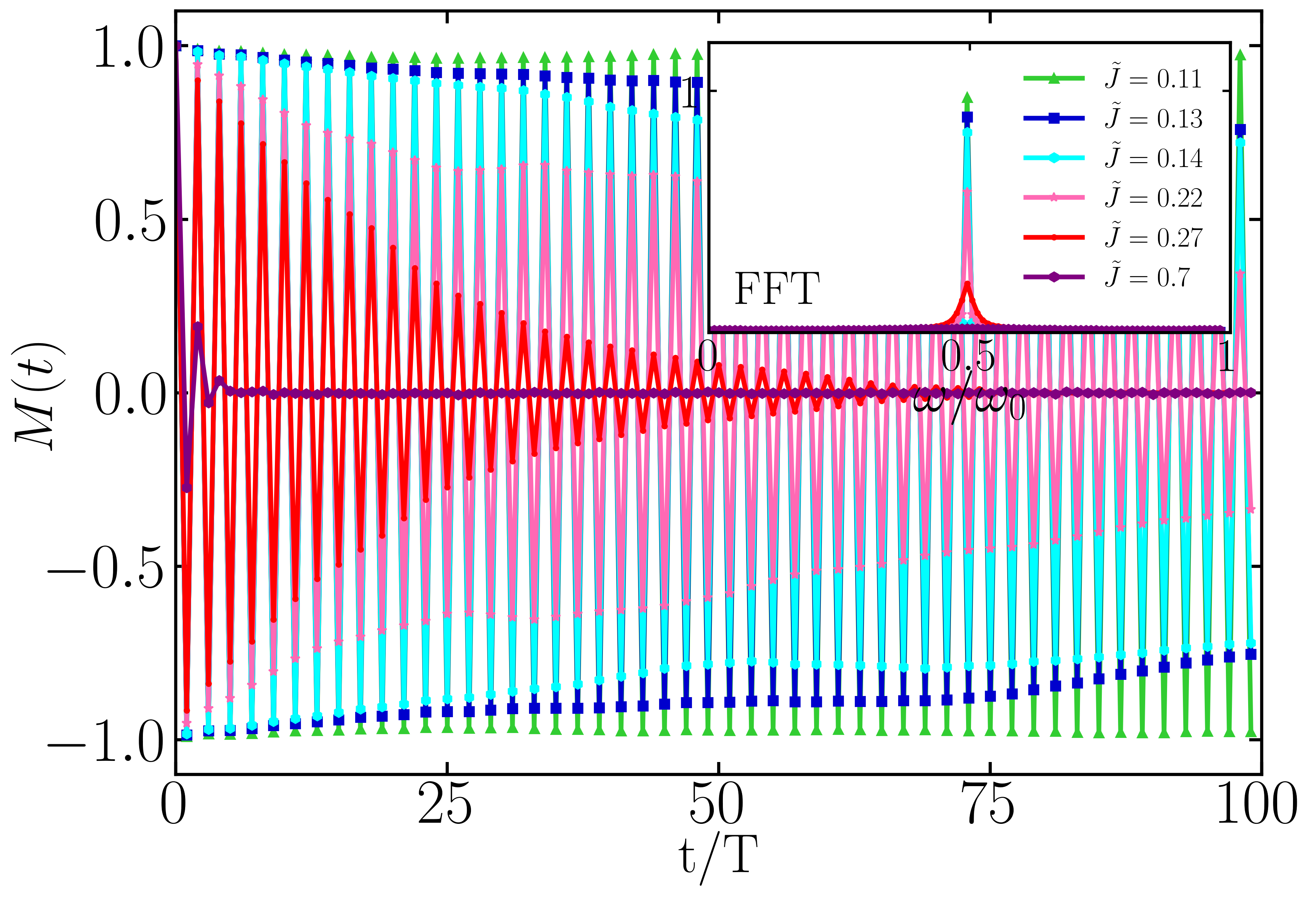}
	\caption{Time evolution of the staggered magnetization $M(t)$ for a dimerized spin chain ($L=18$) with varying interaction strength $\tilde{J}$. For weak $\tilde{J}$ (e.g., $\tilde{J}=0.13$), robust subharmonic oscillations persist due to Floquet QMBS protecting the system from thermalization, indicative of the DTC phase. In contrast, strong interactions (e.g., $\tilde{J}=0.7$) lead to rapid decay of oscillations, signaling the melting of DTC order. Parameters are $\lambda t_2=0.9$ and $\tilde{\Delta}=0.11$. The inset shows the FFT of $M(t)$. A prominent peak at $\omega=\omega_0/2$ in the weak $\tilde{J}$ regime confirms stable $2T$-periodic dynamics, which diminishes with increasing $\tilde{J}$, signaling suppression of coherent oscillations.}
	\label{fig:dtc_dynamics}
\end{figure}

To probe this symmetry breaking, we initialize the system in a product dimer state:
\begin{equation}
|\psi_\gamma\rangle =\bigotimes_{j=1}^{L/2}\left(\cos\gamma |\uparrow_{1,j}\downarrow_{2,j}\rangle+\sin\gamma|\downarrow_{2,j}\uparrow_{1,j}\rangle\right),
\end{equation}
where $j$ indexes the dimer location and $\gamma$ parameterizes the entanglement between the two spins within each dimer. In the limit $\gamma=0$, the spins are unentangled, and the state reduces to the product N\'eel state. Using this $\gamma=0$ limit as a baseline, we monitor the stroboscopic evolution of the staggered magnetization:
\begin{equation}
	M(t)=\frac{1}{L}\sum_{j=1}^{L/2}\langle\psi(t)|\hat{m}_j|\psi(t)\rangle,
\end{equation}
where $\hat{m}_j=\sigma^z_{1,j}-\sigma^z_{2,j}$. We analyze the dynamics of $M(t)$ at integer multiples of the driving period ($t=nT$) in Fig.~\ref{fig:dtc_dynamics}.

In the weak-$\tilde{J}$ regime, $M(t)$ exhibits persistent subharmonic oscillations. In this regime, the QMBS effectively protect the initial state from the thermalizing bulk, preventing the rapid decay of $M(t)$ and maintaining the staggered order. These long-lived, period-doubled oscillations are a definitive signature of the DTC phase. However, as $\tilde{J}$ increases, the system enters a thermalizing regime. The enhanced interaction facilitates extensive mixing across the full Hilbert space, which destroys the scar subspace and leads to the rapid loss of local memory of the initial Néel state. Consequently, the subharmonic oscillations in $M(t)$ decay rapidly, signifying the melting of the DTC order. 

In this context, the lifetime of the DTC response is
characterized by the decay of the oscillation amplitude. For small
$\tilde{J}$, the subharmonic oscillations remain robust, with their amplitude
exhibiting only weak attenuation over many driving cycles. By contrast, for
larger $\tilde{J}$, the envelope of $M(t)$ is suppressed rapidly toward zero,
indicating a substantially shorter-lived DTC response and the onset of
efficient thermalization. Moreover, since $\tilde{J} = J/\lambda$, this
suppression of the DTC order at larger $\tilde{J}$ implies that decreasing
$\lambda$ destroys the DTC phase, whereas increasing $\lambda$ enhances its
stability.

We now extend our analysis to initial states with non-zero intra-dimer entanglement, $|\psi_\gamma\rangle$, where $\gamma \in (0, \pi/4)$. In this regime, any state is a product of entangled dimers. Notably, for $\gamma=\pi/4$, each dimer is maximally entangled.

Crucially, the use of homogeneous, translationally invariant initial states precludes any trivial quasi-MBL phenomena. Unlike inhomogeneous initial configurations, which can induce non-ergodic dynamics through self-generated disorder, our choice of $|\psi_\gamma\rangle$ ensures that any observed persistent non-thermal behavior is unambiguously attributed to the system's intrinsic dynamics—specifically, the presence of QMBS.

\begin{figure}[htbp]
	\centering
	\includegraphics[width=0.45 \textwidth]{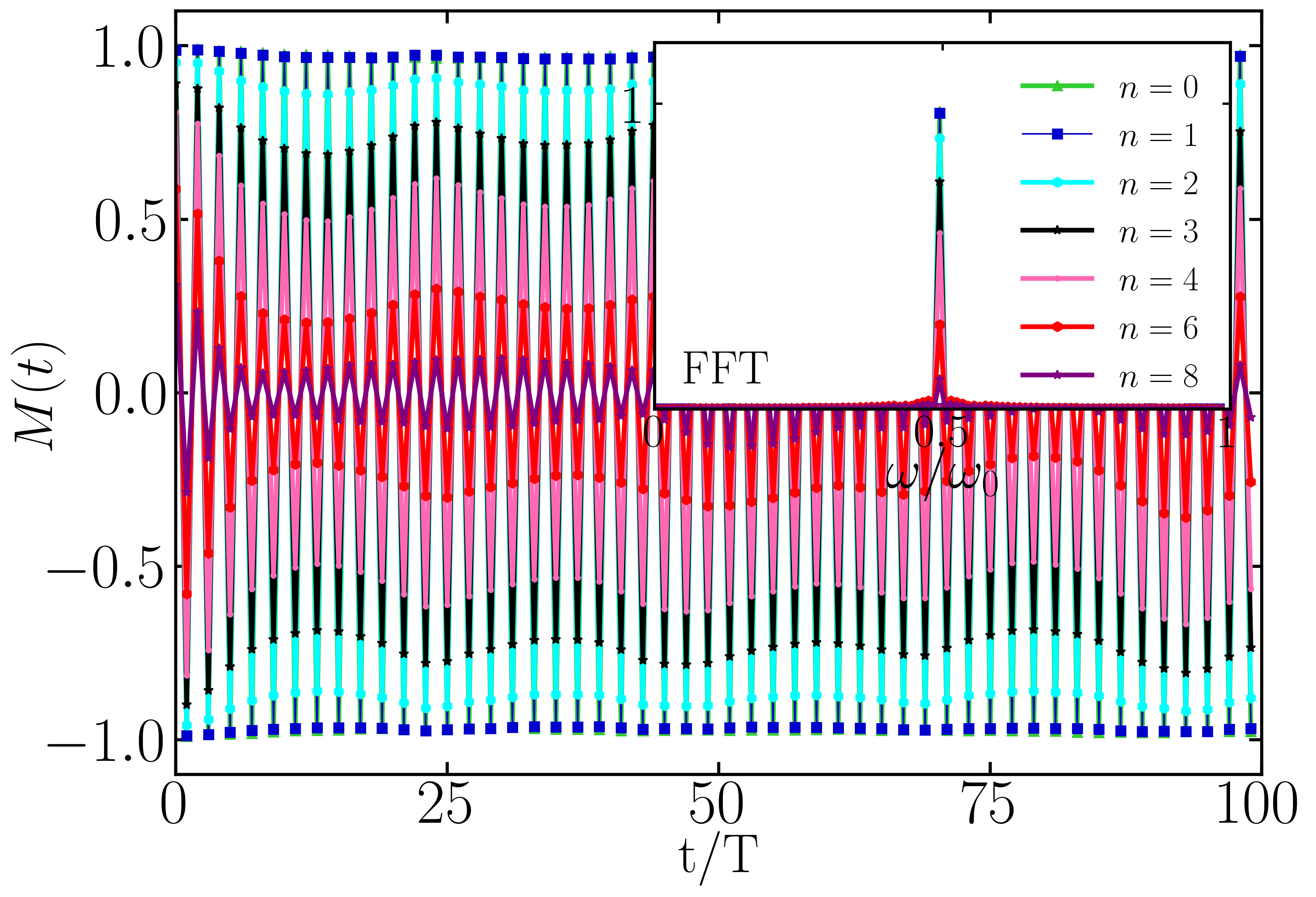}
	\caption{Time evolution of $M(t)$ for initial states $|\psi_\gamma\rangle$ with $\gamma = n\pi/40$ ($n=0, 1, 2, 3, \dots$). As $n$ increases (deviating from the N\'eel state at $n=0$), the initial amplitude of $M(t)$ decreases, but the robust period-doubled oscillations persist without decay. The inset displays the FFT, confirming a strictly locked subharmonic peak at $\omega=\omega_0/2$ for all tested $\gamma$, demonstrating the strong robustness and subharmonic rigidity of the DTC phase against initial state imperfections.}
	\label{fig:magnetization_gamma}
\end{figure}

We monitored the time evolution of the staggered magnetization $M(t)$ from these imperfect initial states. Remarkably, even with initial state deformations, the system continues to exhibit robust period-doubling oscillations that persist for anomalously long times within the scarred regime. While the initial amplitude of $M(t)$ is reduced due to the projection factor from $\gamma$, the underlying frequency and the anomalous lifetime of the subharmonic response remain largely unaffected (see Fig.~\ref{fig:magnetization_gamma}). This resilience indicates that the stabilizing influence of the scarred subspace extends beyond the exact Néel state to a broader neighborhood in the Hilbert space, confirming that the observed DTC order is a robust dynamical phase, not a fine-tuned artifact.

\begin{figure}[htbp]
		\centering
		\includegraphics[width=0.45 \textwidth]{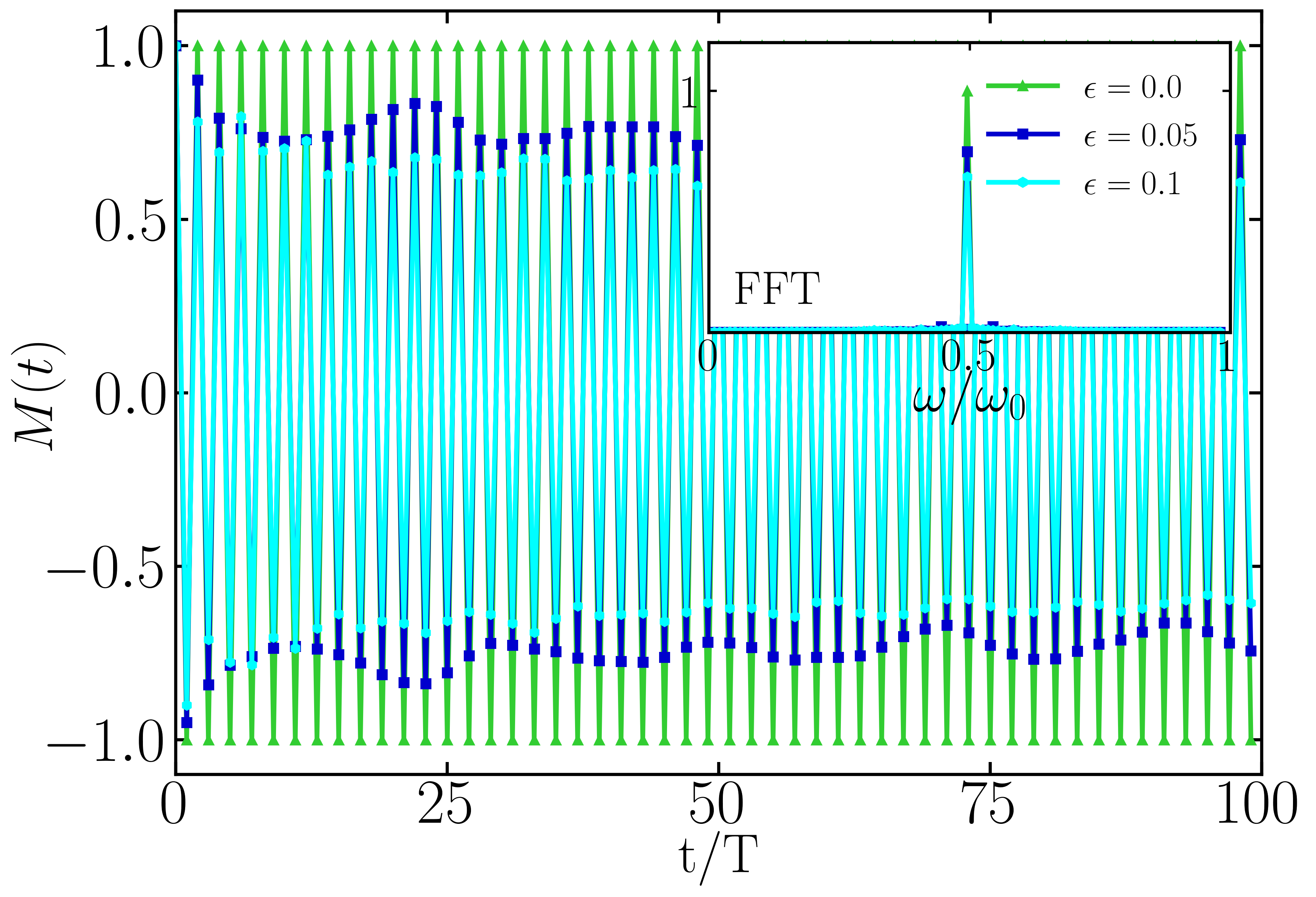}
		\caption{Robustness of the period-doubled oscillations against driving
    imperfections for $\lambda t_2 = 0.9$, $\tilde{\Delta} = 0.11$, and
    $\tilde{J} = 0.11$. The main panel shows the time evolution of the staggered
    magnetization $M(t)$ over $100$ driving cycles for the unperturbed
    ($\epsilon=0$) and perturbed ($\epsilon = 0.05,~ 0.1$) regimes. The
    inset displays the corresponding FFT, demonstrating that the
    dominant frequency remains rigidly locked at the subharmonic value
    $\omega/\omega_0 = 0.5$, signaling stable discrete time-translation
    symmetry breaking.}
		\label{fig:magnetization_imper}
	\end{figure}

Figure~\ref{fig:magnetization_imper} illustrates the time evolution of $M(t)$ together with its Fourier spectrum (inset) under
imperfect rotations parameterized by $\epsilon$, introduced into the driving
sequence as $J t_1 = \pi/4 - \epsilon$. For $\epsilon = 0$, the driving
implements perfect $\pi$-rotations and $M(t)$ displays exact period-doubled
oscillations. Upon adding the perturbation ($\epsilon = 0.05,~ 0.1$), the
time-domain signal develops amplitude modulation and beating; nevertheless, the
subharmonic oscillations persist. This rigidity is confirmed by the FFT spectra
in the inset, where the dominant spectral peak remains locked at
$\omega/\omega_0 = 0.5$ for all values of $\epsilon$ considered. The stability
of this subharmonic peak against such imperfections is a hallmark signature of
the discrete time-translation symmetry breaking that characterizes the SDTC
phase.

We also investigate the evolution of the temporal autocorrelation function, which provides a sensitive probe of dynamical stability and memory retention in driven many-body systems and effectively serves as an order parameter for the DTC phase \cite{marripour2026}. At stroboscopic times $t=nT$, it is defined as
\begin{equation}
	C(nT)=\frac{1}{L}\sum_{j=1}^{L}(-1)^{n}\big\langle \hat{m}_j(nT)\hat{m}_j(0)\big\rangle,
	\label{eq:autocorrelation}
\end{equation}
where $n$ denotes the stroboscopic step. This quantity measures the correlation of the staggered magnetization on a given dimer between the initial time and the later time $nT$. Since the autocorrelation alternates in sign at successive stroboscopic times, the factor $(-1)^n$ compensates for this sign change and renders the behavior smooth from one cycle to the next.

\begin{figure}[htpb]
	\centering
	\includegraphics[width=0.94\linewidth]{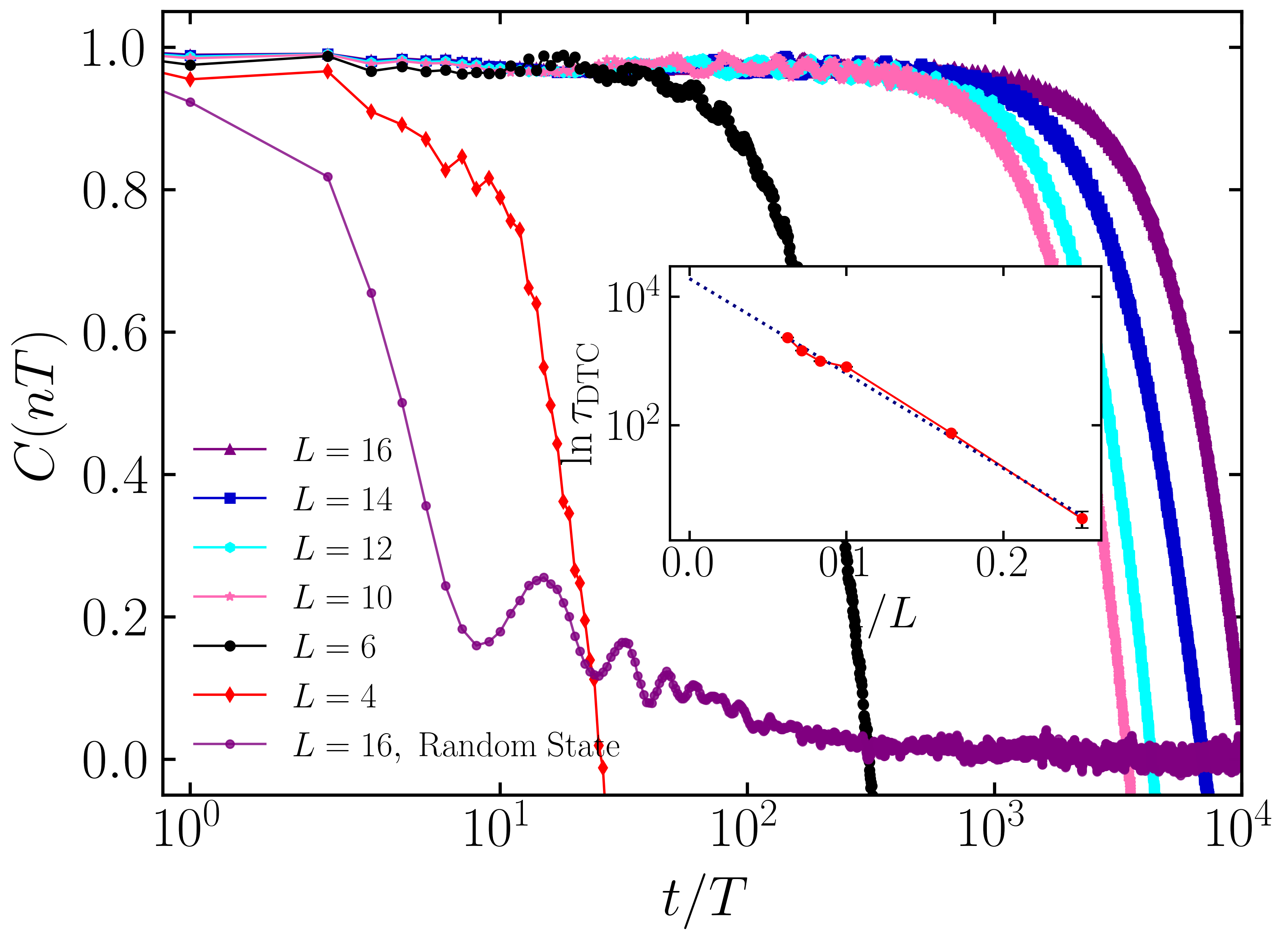}
	\caption{Temporal autocorrelation function $C(nT)$ illustrating the robust DTC phase. With parameters $\lambda t_2=0.9$, $\tilde{\Delta}=0.11$, and $\tilde{J}=0.11$, persistent subharmonic oscillations with period $2T$ are observed for N\'eel-state initialization. The stability of the oscillation amplitude across different system sizes $L$ confirms the robustness of the DTC phase. In contrast, the random initial state exhibits rapid decay, highlighting that the response is highly sensitive to the initial state's entropy. The inset shows the DTC lifetime ($\ln \tau_{\text{DTC}}$) as a function of the inverse system size ($1/L$).}
	\label{fig:autocorrelation}
\end{figure}

As shown in Fig.~\ref{fig:autocorrelation}, we examine $C(nT)$ for different system sizes $L$. In the robust DTC regime, the autocorrelation function exhibits persistent, undamped subharmonic oscillations with period $2T$. Importantly, the finite-size analysis shows that as $L$ increases, the amplitude of these oscillations does not decay, but instead approaches a stable value. This provides strong evidence that the observed DTC order is not a transient finite-size effect, but remains robust as the system size grows.

To investigate the role of the initial state, we also compute the autocorrelation function for a random mixed state corresponding to the infinite-temperature limit ($T_i=\infty$), shown by the purple curve in Fig.~\ref{fig:autocorrelation}. In stark contrast to the N\'eel-state initialization, the autocorrelation for the random state rapidly decays toward zero and fails to develop subharmonic oscillations. This fast loss of memory indicates rapid thermalization and the absence of stable DTC order. Conversely, in the ergodic thermalized regime, $C(nT)$ decays to zero regardless of $L$, signifying a complete loss of initial-state information. These results demonstrate that the observed time-crystalline response is not a generic feature of high-entropy states, but is uniquely associated with the nonergodic dynamical regime accessed from the N\'eel state.

To probe the rigidity of the DTC order, we monitor the stroboscopic evolution of the spin autocorrelation function $C(nT)$ (Fig.~\ref{fig:autocorrelation}). Exact diagonalization up to $L=16$ reveals that for scarred initial states, $C(nT)$ maintains a robust plateau near $1$ before eventually decaying, providing direct evidence of discrete time-translation symmetry breaking. The lifetime of this order, $\tau_{\text{DTC}}$, defined as the duration of the $C(nT) \approx 1$ plateau, increases systematically with system size, extending from $t/T \sim 10$ ($L=4$) to $t/T > 10^4$ ($L=16$). The inset of Fig.~\ref{fig:autocorrelation} shows $\tau_{\text{DTC}}$ versus $1/L$; the observed linear behavior indicates that the lifetime increases exponentially with system size as $\tau_{\text{DTC}} \propto e^{-\alpha/L}$ (with $\alpha > 0$).

Unlike MBL-stabilized DTCs, which prevent thermalization indefinitely, the stability of scar-induced DTCs depends on whether the underlying QMBS are exact or approximate \cite{lin2019,serbyn2021,kolb2023,shiraishi2017}. In models hosting exact scars, the scar states form a decoupled, invariant subspace of the Floquet propagator, allowing oscillations to persist indefinitely in the thermodynamic limit \cite{shiraishi2017}. However, our model analogous to the PXP model \cite{serbyn2021,kolb2023,hudomal2022,Kerschbaumer2025}, features approximate scars. Here, the target initial state is not an exact Floquet eigenstate but exhibits high overlap with a specific subset of the spectrum. Because the dynamical symmetry is imperfect, these states hybridize with the surrounding thermalizing continuum. As $L$ increases, the exponential growth of the density of states facilitates a gradual leakage of quantum information from the initial state into the thermal bath.  

This leakage is, however, slow enough that the dynamical memory of the N\'eel
state remains well protected up to the system sizes accessible to our exact
diagonalization. Indeed, the growth of the DTC lifetime with $L$ (Fig.~\ref{fig:autocorrelation})
is fully compatible with the simultaneous flow of the bulk level-spacing statistics
toward the GOE limit (Figs.~\ref{fig:ps_size} and~\ref{fig:R_Vs_J}): while those
statistics reflect the thermalization of the vast majority of eigenstates in the
dense continuum, the lifetime specifically probes the scarred subspace, which
remains robustly isolated on the length scales we can access. Consequently, the
macroscopic time-crystalline oscillations continue to lengthen with $L$.

We therefore expect $\tau_{\text{DTC}}$ to saturate rather than diverge
with $L$: by analogy with other approximate-scar models, the protection is
eventually overcome by the scar--continuum hybridization, whose strength grows
with the exponentially increasing density of states. The resulting saturation
at a critical size $L_c$ ($\tau_{\text{DTC}} \to \tau_{\text{sat}}$ for
$L \ge L_c$), lying beyond our current numerical reach, marks the observed SDTC
as a long-lived metastable (prethermal) regime rather than a strictly stable
thermodynamic phase.

To further characterize the stability of these nonergodic dynamics, we evaluate the state fidelity, which quantifies the overlap between the time-evolved state and the initial configuration:
\begin{equation}
	F(t)=\left|\langle \psi(0)|\psi(t)\rangle\right|^{2},
	\label{eq:fidelity}
\end{equation}
where $|\psi(0)\rangle$ is the initial state and $|\psi(t)\rangle=U(t)|\psi(0)\rangle$ is the state at time $t$. While $F(t) \approx 1$ indicates high memory retention, $F(t) \approx 0$ signifies a substantial departure from the initial state.

\begin{figure}[htpb]
	\centering
	\includegraphics[width=1.0\linewidth]{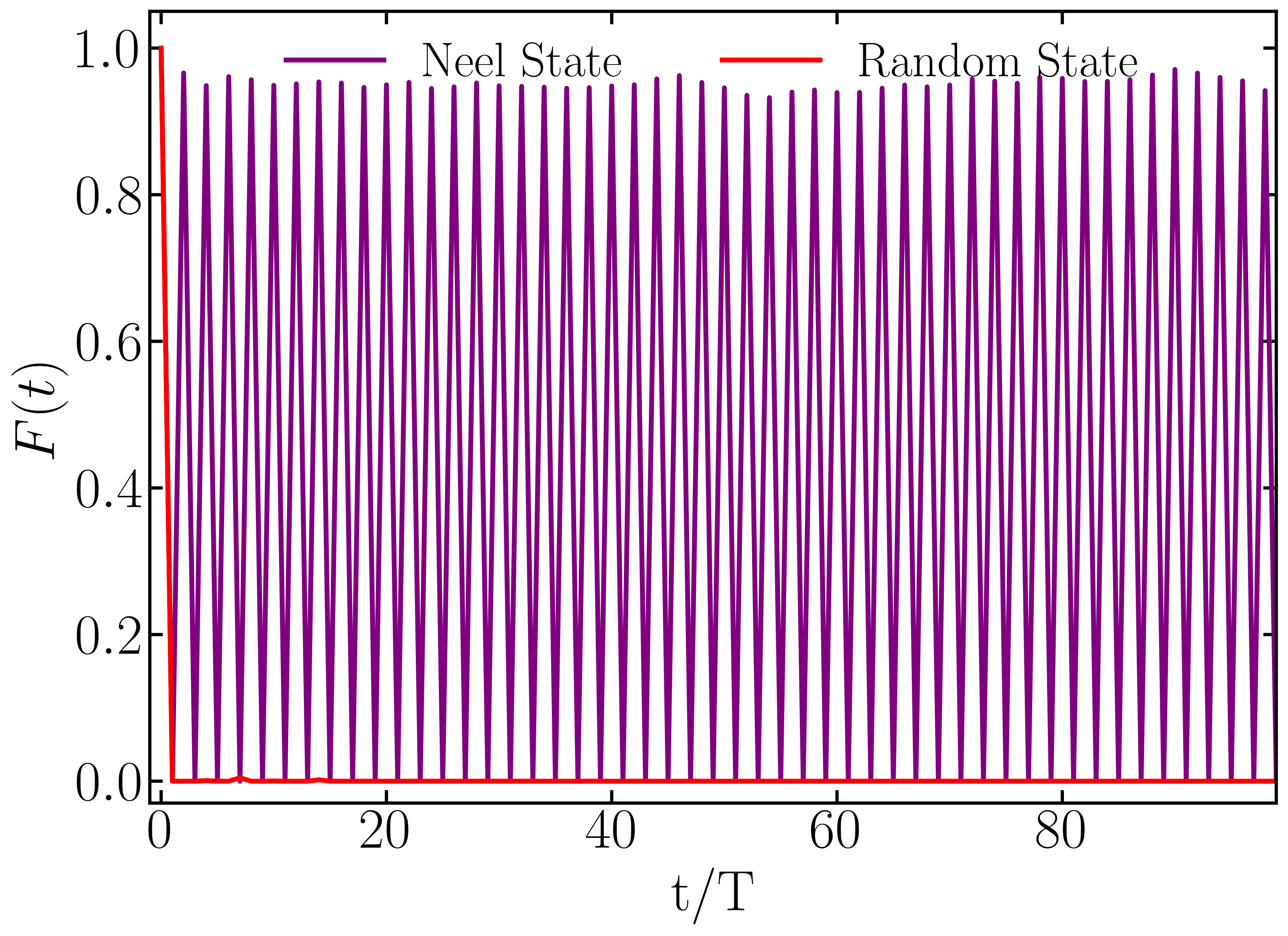}
	\caption{Time evolution of the fidelity $F(t)$ for the N\'eel state and the random state (red solid curve). The N\'eel state shows persistent fidelity revivals, indicating long-lived memory retention and nonergodic dynamics. The rapid decay of the random state signifies fast thermalization and the absence of stable DTC order.}
\label{fig:fidelity}
\end{figure}

Figure~\ref{fig:fidelity} presents the fidelity dynamics for both initial conditions. For the N\'eel state, the fidelity exhibits robust, long-lived oscillations characterized by periodic revivals. This repeated recovery of overlap demonstrates that the system continuously reconstructs its initial configuration, a hallmark of the nonergodic dynamics underlying the DTC phase. In contrast, the fidelity for the random state (infinite-temperature state) drops precipitously and remains near zero, reflecting efficient exploration of the Hilbert space and rapid thermalization. Together, these observations confirm that the persistent memory and subharmonic coherence observed in the N\'eel state are intrinsic to the DTC phase and are not supported by high-entropy initial states.

\begin{figure}[htbp]
	\centering
	\includegraphics[width=0.46\textwidth]{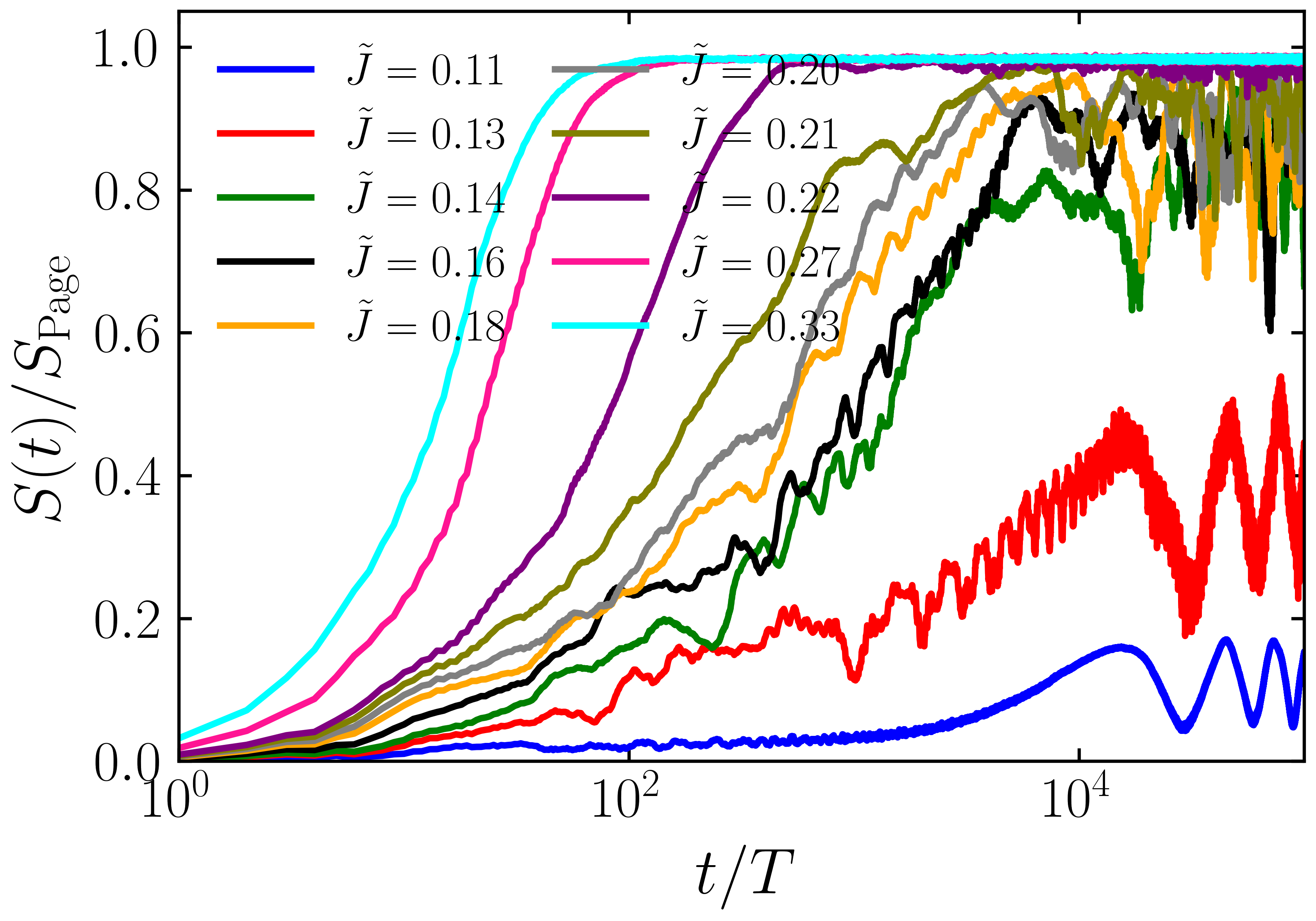}
	\caption{Time evolution of the half-chain entanglement entropy $S(t)/S_{\text{Page}}$ for various coupling strengths $\tilde{J}$, with fixed parameters $\lambda t_2=0.9$ and $\tilde{\Delta}=0.11$. The dynamics transition from a non-ergodic regime (characterized by oscillations and memory preservation) to an ergodic regime (characterized by fast thermalization and saturation at the Page limit). Irregular features in the intermediate regime may stem from finite-size effects or proximity to the phase transition.}
	\label{fig:entropy}
\end{figure}

The time evolution of the EE is illustrated in Fig.~\ref{fig:entropy}, revealing two distinct dynamical regimes governed by the strength $\tilde{J}$.

In the regime ($0<\tilde{J}<0.22$), the system evades rapid thermalization. The EE dynamics are characterized by pronounced, persistent oscillations interspersed with sudden, deep dips. These features are hallmarks of QMBS, manifesting as periodic quantum revivals where the system repeatedly returns toward its initial state, thereby preserving local information.

Conversely, in the regime $\tilde{J}\ge 0.22$, the system exhibits rapid entanglement growth, quickly saturating to the Page limit. In the long-time limit ($t/T>10^3$), the entropy remains flat due to complete thermalization.

%---------------------------------------------------------------------------------------------
\section{Conclusions and Outlook}
\label{sec:conclusion}

In this study, we investigated the non-equilibrium dynamics of a periodically driven dimerized spin chain, demonstrating that the interplay between quantum many-body scars (QMBS) and discrete time-translation symmetry breaking facilitates a robust scarred discrete time crystal (SDTC). Working within the zero-magnetization sector to eliminate statistical mixing, we identified a regime of weak inter-dimer coupling characterized by semi-Poissonian level statistics, weak ETH violation, and the emergence of low-entropy eigenstates (scars) with high overlap with the N\'eel state. Our results establish that this scarred subspace supports long-lived $2T$-periodic oscillations in staggered magnetization and autocorrelation. While SDTC stability is enhanced by optimal inter-dimer interactions ($\lambda t_2$), which suppress phase slips, the system eventually collapses into a thermalized phase when the integrability-breaking parameter $\tilde{J}$ exceeds $\simeq 0.22$, as evidenced by Wigner-Dyson spectral statistics and Page-value entanglement. 
Notably, while our finite-size scaling demonstrates a growing lifetime for the system sizes accessible to exact diagonalization, by analogy with other approximate-scar models, we expect that the eventual hybridization with the thermal continuum will limit this growth at larger sizes. This suggests that the SDTC is a remarkably long-lived prethermal phenomenon rather than a strictly stable thermodynamic phase.

In summary, we have shown that dimerized geometries integrated with Floquet driving offer a robust framework for suppressing thermalization and preserving quantum information, circumventing the conventional requirement for strong disorder. Building on these results, future investigations could prioritize the development of an analytical approach to quantify scar lifetime dependencies on the microscopic parameters of the dimerized chain. Furthermore, extending this framework to quasi-periodically driven systems offers a promising route toward the discovery of scarred time quasicrystals. Finally, the experimental implementation of our model in programmable platforms, such as Rydberg atom arrays or superconducting qubits, remains a critical objective, providing a pathway to empirically validate the synergy between many-body scarring and discrete time-crystalline order.

% ==========================================================
% APPENDIX
% ==========================================================

\appendix
\section{Level spacing statistics and crossover regimes}
\label{app:level_statistics}

\begin{figure}[htbp]
	\centering
	\includegraphics[width=0.48\textwidth]{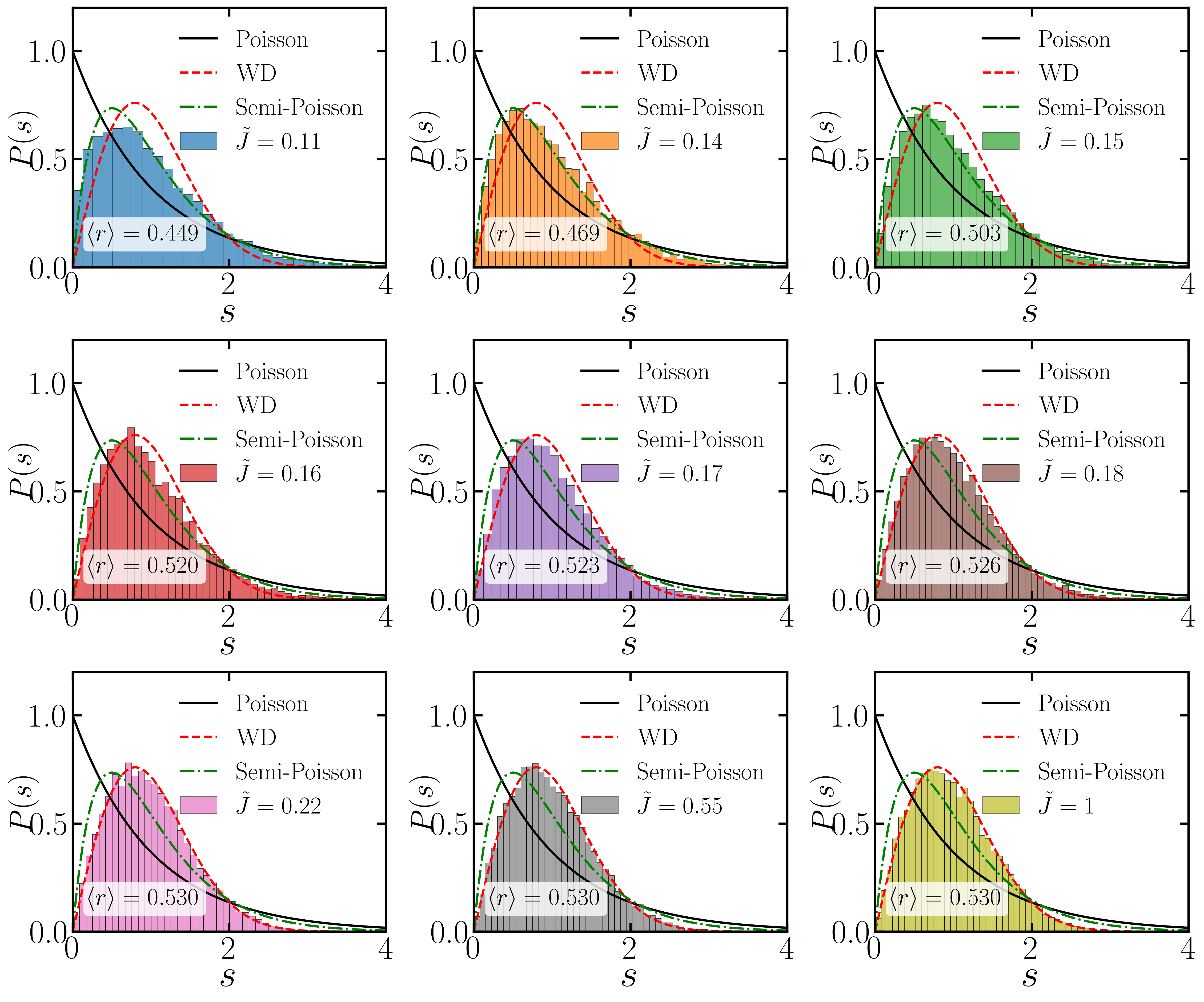}
	\caption{Level-spacing distribution $P(s)$ for different values of $\tilde{J}$. The panels illustrate the crossover of the system's statistics from a semi-Poisson distribution at small $\tilde{J}$ (indicative of QMBS) to a WD distribution at larger $\tilde{J}$ (thermalized phase).}
	\label{fig:my_wide_figure}
\end{figure}

A systematic analysis of the level-spacing distribution $P(s)$ reveals three distinct physical regimes in our finite-size dimerized spin model. For fixed parameters $\lambda t_2 = 0.9$, $\tilde{h}=1$, and $\tilde{\Delta} = 0.11$, tuning the coupling strength $\tilde{J}$ drives a crossover between these regimes, as illustrated in Fig.~\ref{fig:my_wide_figure}:

The regimes are characterized as follows:

\begin{itemize}
    \item {Scar-dominated regime ($0 < \tilde{J} \leq 0.15$):} The signature of QMBS is highly prominent. Due to the presence of these non-ergodic states, the $P(s)$ distribution closely follows semi-Poisson statistics.
    
    \item {Crossover region ($0.15 < \tilde{J} < 0.22$):} This interval signifies a transition where the dominant role of quantum scars gradually diminishes, though their residual effects persist. Consequently, the $P(s)$ distribution exhibits a departure from semi-Poisson statistics, shifting steadily toward the WD distribution.
    
    \item {Ergodic/Thermalized regime ($\tilde{J} \geq 0.22$):} The system becomes fully thermalized. In this regime, the level spacing statistics show excellent agreement with the WD prediction, as evidenced by the $\tilde{J}=0.55$ and $\tilde{J}=1$ panels, where numerical histograms align closely with the theoretical WD curves.
\end{itemize}

\begin{thebibliography}{99}

			
	\bibitem{wang2024ETH}
	D.-Z. Wang, H. Zhu, J. Cui, J. Argüello-Luengo, M. Lewenstein, G.-F. Zhang, and P. Sierant,
	"Eigenstate thermalization and its breakdown in quantum spin chains with inhomogeneous interactions," \textit{Physical Review B} {\bf 109}, 045139 (2024).

	\bibitem{srednicki1994}
	M. Srednicki, "Chaos and quantum thermalization," \textit{Physical Review E} {\bf 50}, 888--901 (1994).

	\bibitem{rigol2008}
	M. Rigol, V. Dunjko, and M. Olshanii,"Thermalization and its mechanism for generic isolated quantum systems," \textit{Nature} {\bf 452}, 854--858 (2008).

    \bibitem{dunajski2012}
    M. Dunajski,"Integrable Systems,"Department of Applied Mathematics and Theoretical Physics, University of Cambridge, Wilberforce Road, Cambridge CB3 0WA, UK (2012).

    \bibitem{basko2006}
    D. M. Basko, I. L. Aleiner, and B. L. Altshuler,
    "Metal--insulator transition in a weakly interacting many-electron system with localized single-particle states," \textit{Annals of Physics} {\bf 321}, 1126--1205 (2006).
    
	\bibitem{nandkishore2015}
	R. Nandkishore and D. A. Huse,
	"Many-body localization and thermalization in quantum statistical mechanics,"
	\textit{Annual Review of Condensed Matter Physics} {\bf 6}, 15--38 (2015).
	
    \bibitem{kumar2024hilbert}
    A. Kumar, et al., "Hilbert Space Fragmentation and Scar Time Crystallinity in Driven Homogeneous Central Spin Models," \textit{APS March Meeting Abstracts} {\bf 2024}, (2024).
    \bibitem{Langlett}
    C. M. Langlett and S. Xu, "Hilbert space fragmentation and exact scars of generalized Fredkin spin chains," \textit{Physical Review B} \textbf{103}, L220304 (2021).
     \bibitem{Moudgalya1}
    S. Moudgalya, B. A. Bernevig, and N. Regnault, "Quantum many-body scars and Hilbert space fragmentation: A review of exact results," \textit{Reports on Progress in Physics} \textbf{85}, 086501 (2022).
    	%%%%%%%%%%%%%%%%%%%%%%%%%%%%%%%%%%%%%%%%%%%%%%%%%%%%%%%%%prethermal
    \bibitem{B. Bauer} 
    D. V. Else, B. Bauer, and C. Nayak, "Prethermal phases of matter protected by time-translation symmetry," \textit{Physical Review X} \textbf{7}, 011026 (2017).
    \bibitem{K. Mallayya} 
    K. Mallayya, M. Rigol, and W. De Roeck, "Prethermalization and thermalization in isolated quantum systems," \textit{Physical Review X} \textbf{9}, 021027 (2019).
    
    \bibitem{D. V. Else} D. V. Else, W. W. Ho, and P. T. Dumitrescu, Long-lived Interacting Phases of Matter Protected by Multiple Time-Translation Symmetries in Quasiperiodically Driven Systems, \textit{Physical Review X} \textbf{10}, 021032 (2020).
    \bibitem{G. He} 
    G. He et al., "Quasi-Floquet prethermalization in a disordered dipolar spin ensemble in diamond," \textit{Physical Review Letters} \textbf{131}, 130401 (2023).
    \bibitem{S. A. Weidinger} 
    S. A. Weidinger and M. Knap, "Floquet prethermalization and regimes of heating in a periodically driven, interacting quantum system," \textit{Scientific Reports} \textbf{7}, 45382 (2017).
    \bibitem{E. Canovi} 
    E. Canovi, M. Kollar, and M. Eckstein, "Stroboscopic prethermalization in weakly interacting periodically driven systems," \textit{Physical Review E} \textbf{93}, 012130 (2016).
    \bibitem{M. Bukov} 
    M. Bukov, S. Gopalakrishnan, M. Knap, and E. Demler, "Prethermal Floquet steady states and instabilities in the periodically driven, weakly interacting Bose-Hubbard model," \textit{Physical Review Letters} \textbf{115}, 205301 (2015).
    
    \bibitem{Kyprianidis}
    A. Kyprianidis, F. Machado, W. Morong, P. Becker, K. S. Collins, D. V. Else, ..., and C. Monroe, "Observation of a prethermal discrete time crystal," \textit{Science} \textbf{372}, 1192--1196 (2021).
    \bibitem{Das Sarma}
    D. Vu and S. Das Sarma, "Dissipative prethermal discrete time crystal," \textit{Physical Review Letters} \textbf{130}, 130401 (2023)
    \bibitem{Zeng}
    T. S. Zeng and D. N. Sheng, "Prethermal time crystals in a one-dimensional periodically driven Floquet system," \textit{Physical Review B} \textbf{96}, 094202 (2017).
    \bibitem{Stasiuk}
    A. Stasiuk and P. Cappellaro, "Observation of a prethermal U(1) discrete time crystal," \textit{Physical Review X} \textbf{13}, 041016 (2023).
    \bibitem{Nandkishore}
    R. Nandkishore, S. Gopalakrishnan, and D. A. Huse, "Spectral features of a many-body-localized system weakly coupled to a bath," \textit{Physical Review B} \textbf{90}, 064203 (2014).
    \bibitem{Lazarides} 
    A. Lazarides, A. Das, and R. Moessner, "Fate of many-body localization under periodic driving," \textit{Physical Review Letters} \textbf{115}, 030402 (2015).
    \bibitem{Kjall}
    J. A. Kjäll, J. H. Bardarson, and F. Pollmann, "Many-body localization in a disordered quantum Ising chain," \textit{Physical Review Letters} \textbf{113}, 107204 (2014).
    \bibitem{huse}
    Arijeet Pal and David A. Huse, “Many-body localization
    phase transition,” \textit{Physical Review B} \textbf{82}, 174411 (2010).
    \bibitem{Bordia}
    P. Bordia, H. Lüschen, U. Schneider, M. Knap, and I. Bloch, "Periodically driving a many-body localized quantum system," \textit{Nature Physics} \textbf{13}, 460--464 (2017).
    
    \bibitem{Oganesyan}
    V. Oganesyan and D. A. Huse, "Localization of interacting fermions at high temperature," \textit{Physical Review B} \textbf{75}, 155111 (2007).
    \bibitem{Keyserlingk}
    C. W. von Keyserlingk and S. L. Sondhi, "Phase structure of one-dimensional interacting Floquet systems. II. Symmetry-broken phases," \textit{Physical Review B} \textbf{93}, 245146 (2016).
    \bibitem{Johri}
    S. Johri, R. Nandkishore, and R. N. Bhatt, "Many-body localization in imperfectly isolated quantum systems," \textit{Physical Review Letters} \textbf{114}, 117401 (2015).
    \bibitem{Huse}
    D. A. Huse, R. Nandkishore, V. Oganesyan, A. Pal, and S. L. Sondhi, "Localization-protected quantum order," \textit{Physical Review B} \textbf{88}, 014206 (2013).
    \bibitem{P. Ponte}
    P. Ponte, Z. Papić, F. Huveneers, and D. A. Abanin, "Many-body localization in periodically driven systems," \textit{Physical Review Letters} \textbf{114}, 140401 (2015).
    
    \bibitem{Igloi2007}
    F. Iglói, R. Juhász, and Z. Zimborás, 
    ``Entanglement entropy of aperiodic quantum spin chains,'' 
    Europhys. Lett. \textbf{79}, 37001 (2007).
    \bibitem{bernien2017}
    H. Bernien et al., "Probing many-body dynamics on a 51-atom quantum simulator,"
    \textit{Nature} {\bf 551}, 579--584 (2017).
    
    \bibitem{moudgalya2018}
    S. Moudgalya, S. Rachel, B. A. Bernevig, and N. Regnault,
    "Exact excited states of nonintegrable models,"
    \textit{Physical Review B} {\bf 98}, 235155 (2018).
    
    \bibitem{kunimi2024}
    M. Kunimi et al.,
    "Proposal for simulating quantum spin models with the Dzyaloshinskii-Moriya interaction using Rydberg atoms and the construction of asymptotic quantum many-body scar states,"
    \textit{Physical Review A} {\bf 110}, 043312 (2024).
    
    \bibitem{bluvstein2021}
    D. Bluvstein et al., "Controlling quantum many-body dynamics in driven Rydberg atom arrays,"
    \textit{Science} {\bf 371}, 1355--1359 (2021).
    \bibitem{su2022}
    G. X. Su, H. Sun, A. Hudomal, J. Y. Desaules, Z. Zhou, B. Yang, et al.,
    "Observation of unconventional many-body scarring in a quantum simulator,"
    \textit{Nature Physics} (2022).
	
	\bibitem{oka2019}
	T. Oka and S. Kitamura,
	``Floquet engineering of quantum materials,''
	\textit{Annual Review of Condensed Matter Physics} {\bf 10}, 387--408 (2019).
	
	\bibitem{khemani2016}
	V. Khemani, A. Lazarides, R. Moessner, and S. L. Sondhi,
	``Phase structure of driven quantum systems,''
	\textit{Physical Review Letters} {\bf 116}, 250401 (2016).
	
		\bibitem{else2016}
	D. V. Else, B. Bauer, and C. Nayak,
	``Floquet time crystals,''
	\textit{Physical Review Letters} {\bf 117}, 090402 (2016).
	
	\bibitem{marripour2025}
	D. Marripour and J. Abouie,
	"From time crystals to time quasicrystals: Exploring quasiperiodic phases in transverse field Ising chains,''\textit{Physical Review B} {\bf 112}, 174307 (2025).
	
	\bibitem{Das2026}
	G. Das, S. Saha, and R. Bhattacharyya, "Environment-assisted discrete time crystals in noninteracting quantum systems," \textit{Physical Review A} \textbf{113}, 042216 (2026).
	
	\bibitem{marripour2026}
	D. Marripour and J. Abouie,
	"Emergence of prethermal time-quasicrystalline order in a quasiperiodically driven noninteracting spin chain", \textit{Physical Review B} {\bf 113}, 214318 (2026).

	\bibitem{yao2017}
	N. Y. Yao et al.,
	"Discrete time crystals: rigidity, criticality, and realizations,''
	\textit{Physical Review Letters} {\bf 118}, 030401 (2017).

	\bibitem{zhang2017}
	J. Zhang et al.,
	"Observation of a discrete time crystal,''
	\textit{Nature} {\bf 543}, 217--220 (2017).
	
	\bibitem{Huang2018}
	Huang, Biao, Ying-Hai Wu, and W. Vincent Liu. "Clean Floquet time crystals: models and realizations in cold atoms", \textit{Physical Review Letters} {\bf 120}, 110603 (2018).
	
	\bibitem{H. Yar} 
	H. Yarloo, A. Emami Kopaei, and A. Langari, "Homogeneous Floquet time crystal from weak ergodicity breaking", \textit{Physical Review B} {\bf 102}, 224309 (2020).
	
	\bibitem{maskara2021}
	N. Maskara, A. A. Michailidis, W. W. Ho, D. Bluvstein, S. Choi, M. D. Lukin, and M. Serbyn,
	``Discrete time-crystalline order enabled by quantum many-body scars: Entanglement steering via periodic driving'',
	\textit{Physical Review Letters} {\bf 127}, 090602 (2021).
	
	\bibitem{sugiura2021}
	S. Sugiura, T. Kuwahara, and K. Saito,
	"Many-body scar state intrinsic to periodically driven system: Rigorous results,''
	\textit{Physical Review Research} {\bf 3}, L012010 (2021).
	
	
	\bibitem{ho2019}
	W. W. Ho, S. Choi, H. Pichler, and M. D. Lukin,
	``Periodic dynamics in Rydberg atom arrays,''
	\textit{Physical Review Letters} {\bf 122}, 040603 (2019).
		\bibitem{hudomal2022}
	A. Hudomal, J. Y. Desaules, B. Mukherjee, G. X. Su, J. C. Halimeh, and Z. Papić,
	``Driving quantum many-body scars in the PXP model,''
	\textit{Physical Review B} {\bf 106}, 104302 (2022).
	
	
	\bibitem{turner2018}
	C. J. Turner, A. A. Michailidis, D. A. Abanin, M. Serbyn, and Z. Papić,
	``Weak ergodicity breaking from quantum many-body scars,''
	\textit{Nature Physics} {\bf 14}, 745 (2018).
	
	\bibitem{mukherjee2020}
	B. Mukherjee, S. Sen, D. Sinha, and K. Sengupta,
	``Collapses and revivals of quantum many-body scars via Floquet engineering,''
	\textit{Physical Review Research} {\bf 2}, 043317 (2020).
	\bibitem{Atas2013}
	Y. Y. Atas, E. Bogomolny, O. Giraud, and G. Roux, "Distribution of the ratio of consecutive level spacings in random matrix ensembles", \textit{Physical Review Letters} \textbf{110}, 084101 (2013).
	
	\bibitem{Schmit1999}
	Bogomolny, E. B., Gerland, U., Schmit, C " Models of intermediate spectral statistics". \textit{Physical Review E} \textbf{59(2)}, R1315 (1999).
	\bibitem{Page1993}
	D. N. Page, "Average entropy of a subsystem", \textit{Physical Review Letters} \textbf{71}, 1291 (1993).
	
	\bibitem{serbyn2021}
	M. Serbyn, D. A. Abanin, and Z. Papi{\'c},
	"Quantum many-body scars and weak breaking of ergodicity," \textit{Nature Physics} {\bf 17}, 675--685 (2021).
	\bibitem{kolb2023}
	P. Kolb and K. Pakrouski,
	"Stability of the many-body scars in fermionic spin-1/2 models," \textit{PRX Quantum} {\bf 4}, 040348 (2023).
	
	\bibitem{shiraishi2017}
	N. Shiraishi and T. Mori,
	"Systematic construction of counterexamples to the eigenstate thermalization hypothesis," \textit{Physical Review Letters} {\bf 119}, 030601 (2017).
	
	\bibitem{lin2019}
	C.-J. Lin and O. I. Motrunich, "Exact quantum many-body scar states in the Rydberg-blockaded atom chain," \textit{Physical Review Letters} {\bf 122}, 173401 (2019).

	\bibitem{Kerschbaumer2025}
	A. Kerschbaumer, et al., “Quantum many-body scars beyond the PXP model in Rydberg simulators,” \textit{Physical Review Letters} {\bf 134}, 160401 (2025).


		
	\end{thebibliography}
\end{document}